\documentclass[usenatbib]{mn2e}

\usepackage{mathptmx}
\usepackage{times}

\usepackage{epsfig}
\usepackage{graphicx}
\usepackage{graphics}
\usepackage{color}
\usepackage{mathrsfs}

\def\ionh{H{\small{II}}}
\def\ionhe1{He{\small{II}}}
\def\ionhe2{He{\small{III}}}

\title{On the role of recombination in common-envelope ejections}

\author[N. Ivanova, S. Justham and Ph.~Podsiadlowski]
  {N. Ivanova,$^{1}$\thanks{E-mail:nata.ivanova@ualberta.ca} S. Justham$^2$ and Ph.~Podsiadlowski$^3$\\ 
$^1${University of Alberta, Dept.\ of Physics, 11322-89 Ave, Edmonton, AB, T6G 2E7, Canada} \\
$^2${National Astronomical Observatories, Chinese Academy of Sciences,
Beijing 100012, China} \\
$^3${Oxford Astrophysics, Denys Wilkinson Building, Keble Road, Oxford, OX1 3RH, United Kingdom}
}

\begin{document}

\maketitle

\begin{abstract} 
The energy budget in common-envelope events (CEEs) is not well
understood, with substantial uncertainty even over to what extent the
recombination energy stored in ionised hydrogen and helium might be
used to help envelope ejection. We investigate the reaction of a
red-giant envelope to heating which mimics limiting cases of energy
input provided by the orbital decay of a binary during a CEE,
specifically during the post-plunge-in phase during which the
spiral-in has been argued to occur on a time-scale longer than
dynamical. We show that the outcome of such a CEE depends less on the
total amount of energy by which the envelope is heated than on
\emph{how rapidly} the energy was transferred to the envelope and on
\emph{where} the envelope was heated.  The envelope always becomes
dynamically unstable \emph{before} receiving net heat energy equal to
the envelope's initial binding energy. We find two types of outcome,
both of which likely lead to at least partial envelope ejection:
``runaway'' solutions in which the expansion of the radius becomes
undeniably dynamical, and superficially ``self-regulated'' solutions,
in which the expansion of the stellar radius stops but a significant
fraction of the envelope becomes formally dynamically unstable. Almost
the entire reservoir of initial helium recombination energy is used
for envelope expansion. Hydrogen recombination is less energetically
useful, but is nonetheless important for the development of the
dynamical instabilities. However, this result requires the companion
to have already plunged deep into the envelope; therefore this release
of recombination energy does not help to explain wide
post-common-envelope orbits.
\end{abstract}

\begin{keywords}
binaries: close --- stars: evolution --- stars: mass-loss
\end{keywords}

\section{Introduction}

A common-envelope event (CEE) is a brief episode in the life of a
binary star during which two stars share an envelope that surrounds
their orbit \citep[][]{Pa76}. For the cases in which the envelope is
successfully ejected, a CEE might be regarded as a temporary merger,
capable of transforming an initially wide  binary system into a compact
binary. This mechanism is thought to be involved in the production of
X-ray binaries, type Ia supernova progenitors and stellar-mass gravitational-wave
sources, and is also important in binary models for the progenitors of
gamma-ray bursts. However, whilst CEEs are vitally important for 
a significant fraction of all binaries including 
the production of a wide variety of energetic stellar exotica, the overall
process is still poorly understood. 

The overwhelming reasons for our difficulty in understanding CEEs are
the complexity of the physical processes involved in each CEE and the
extreme ranges in both time-scales (up to a factor of $10^{10}$) and
length-scales (up to $10^8$) on which those physical processes take
place \citep{Ivanova+2013Review}. For example, it has been shown first in
one-dimensional studies that employed stellar evolution codes
\citep{mmh79,podsi01}, and then later re-confirmed by three-dimensional
studies that used different hydrodynamical codes \citep{Ric08,Passy11},
that for most considered binary configurations the orbital shrinkage,
also known as spiral-in, slows down from evolving on a dynamical
time-scale  -- this phase is referred to as the plunge-in --  to a thermal
time-scale, which is referred to as a ``self-regulated'' spiral-in. At
this slow stage, modern hydrodynamical codes are no longer capable of
treating the physics involved (e.g., convective energy transport), nor
are they able to integrate for the expected duration of the stage whilst
keeping the most important quantities, such as angular momentum and
energy, conserved \citep{Ivanova+2013Review}.

During the self-regulated spiral-in, the presence of the secondary
within the primary star results in energy being deposited into the
envelope of the primary. This heating luminosity is normally expected to
be dominated by dissipation of orbital energy, i.e.\ release of
gravitational potential energy during the in-spiral, although it is
unclear which dissipation mechanism dominates; candidates include
viscous friction in differentially rotating layers, tidal friction, and
spiral shocks.  Another potential source of heating luminosity is
accretion onto the secondary star. Convection is expected to advect the
energy to the surface, and the envelope is expected to adapt its
structure to the new total luminosity.

However, the self-regulated spiral-in can not continue forever, and one
of the possible endings is an eventual ejection of the envelope
\citep[e.g., via delayed dynamical ejection][]{IvanovaThesis,Han+2002}.
Whilst delayed dynamical envelope ejection was reported to take place in
those simulations, there were no clear physical reasons for the
ejection.

The situation of envelope ejection during a CEE may well be very 
analogous to the case of AGB stars, during which stars eject their own
luminous envelopes to form planetary nebulae.  The specific instability
which causes AGB envelope ejection is still debated, but numerous
previous studies exist \citep[see,
e.g.,][]{PZ1968,Kutter+Sparks1972,Sparks+Kutter1972,Kutter+Sparks1974,
Tuchman+1978,Tuchman+1979,Fox+Wood1982,W+W1994,Han+1994,2008ApJ...674L..49S}.

One further similarity between CEEs and AGB stars is that it has been
suggested that energy released from the recombination of ionised plasma
may help to eject the envelope in both cases \citep[see
especially][]{PZ1968,W+W1994,Han+1994,Han+1995,Han+2002,Ivanova+2013Review}.
Whilst evidence was recently presented which suggests that hydrogen
recombination is unlikely to be helpful in unbinding the envelope for a
large fraction of CEE events \citep[since recombination happens very
near to the photosphere of the already ejected
envelope;][]{Ivanova+2013Science}, our physical understanding of envelope
ejection remains highly incomplete.  In addition, since helium
recombination is expected to occur at higher optical depths than
hydrogen recombination, it is particularly plausible that helium
recombination could affect the outcome of a CEE \citep[see the discussion
in \S 3.3.2 of][]{Ivanova+2013Review}.

This paper systematically examines the physics of envelope ejection and
the development of instabilities during CEEs. We adopt a simplified model
in order to try to understand the energy budget and the criteria for
envelope ejection. The main difference between these calculations and
previous work on AGB envelopes is the addition of an artificial heating
term. This term is intended to approximately mimic the energy release
during the spiral-in of the companion star during CEE in a deliberately
simplified way.

We introduce the main quantities that we use for the analysis of
instabilities and of the ejection criterion in \S2 and describe the
numerical method for the stellar heating in \S3. We describe the
outcomes for the two limiting cases we use to heat the model star in \S4
and \S5. In \S6 we analyze the physical processes that affect the
results, including the development of instabilities in the envelope and
potential envelope ejection.

\section{Quantities}

\label{sec:quant}

We first introduce several important integrated properties of matter in
the envelope: the potential energy $E_{\rm pot}$, the internal energy
$E_{\rm int}$, enthalpy $H=E_{\rm int}+ P/V$, recombination energy
$E_{\rm rec}$ (which we define as the energy which is stored in ionised
matter) and kinetic energy $E_{\rm kin}$. In all cases, these are
functions of the mass from which the quantity was integrated to the
surface:

\begin{equation} E_{\rm pot}(m_{\rm bot})=\int_{m_{\rm bot}}^{M}
\frac{Gm}{r}\, dm; \end{equation}

\begin{equation} E_{\rm int}(m_{\rm bot})=\int_{m_{\rm bot}}^{M} u\, dm;
\end{equation}

\begin{equation} H(m_{\rm bot})= E_{\rm int} (m_{\rm bot}) +
\int_{m_{\rm bot}}^{M_*} P/\rho\, dm ; \end{equation}

\begin{equation} E_{\rm rec}(m_{\rm bot})=\int_{m_{\rm bot}}^{M}
\varepsilon_{\rm rec}\, dm ; \end{equation}

\begin{equation} E_{\rm kin}(m_{\rm bot})=\int_{m_{\rm bot}}^{M}
\frac{1}{2} v^2 \, dm . \end{equation}

\noindent Here $M$ is the total mass of the star, $m$ is the local mass
coordinate, $r$ is the radial coordinate, $u$ is the specific internal
energy (no recombination energy is taken into account), $\varepsilon_{\rm rec}$ is the specific recombination energy,
$P$ is the pressure and $\rho$ is the density. The recombination energy
stored in ionised matter can be further described through its three main
components, i.e.\ the energy stored in ionised hydrogen ($E_{\rm
rec}^{\rm H{\small I}}$), singly ionised helium ($E_{\rm rec}^{\rm He{\small{}II}}$) and doubly
ionised helium ($E_{\rm rec}^{\rm He\small{III}}$).

The binding energy of the envelope above $m_{\rm bot}$, as is typically
used for estimates of the CE energy budget, is then

\begin{equation} E_{\rm bind}(m_{\rm bot}) = E_{\rm pot}(m_{\rm
bot})+E_{\rm int}(m_{\rm bot}) \ . \end{equation}

\noindent The common-envelope energy formalism presumes that this amount
of energy must be supplied to eject the envelope, and equates this
requirement with the energy available from orbital decay. However, we
stress that it is not clear what energy reservoir is truly available
\citep{Webbink08,Ivanova+2013Review} nor how much energy is truly
required to expel the envelope and whether that energy is as simply
linked to the envelope binding energy as assumed above \citep{Ivach11}.

An exact determination of the post-CE core is very important for the
energy balance, as the binding energy of the layers closest to the core
can dramatically change the overall envelope binding energy \citep[see,
e.g.,][]{TaurisDewi01,DeloyeTaam10,Ivanova11,Ivanova+2013Review}.
Unfortunately, it is not clear whether the outer boundary of a post-CE
core would be at the inner mass coordinate of the ejected envelope,
$m_{\rm bot}$, as might reasonably be expected  \citep[see, e.g., the
discussion about a possible post-CE thermal time-scale mass transfer
in][]{Ivanova11}.

Dynamical stability is often characterized using the first adiabatic
index $\Gamma_1$ \citep{Ledoux1945}: 

\begin{equation} \Gamma_1=\left (
\frac{\partial \ln P}{\partial \ln \rho}\right )_{\rm ad}  .
\end{equation}

The dynamical stability criterion then depends on the pressure-weighted
volume-averaged value of $\Gamma_{1}$ (\citealt{Ledoux1945}; see also
\citealt{Stothers1999}): 

\begin{equation} \langle \Gamma_1 (m_{\rm bot)
}\rangle = \frac{\int_{m_{\rm bot}}^{M_*} \Gamma_1 P dV }{\int_{m_{\rm
bot}}^{M_*} P dV} \ , \end{equation}

\noindent such that, if $\langle \Gamma_1 (0) \rangle < 4/3$, the whole
star is dynamically unstable. \cite{Lobel2001} argued that the envelope
in cool giants becomes unstable if $\langle \Gamma_1 (m_{\rm env})
\rangle< 4/3$ where $m_{\rm env}$ is the bottom of the envelope.

However, it is not clear where $m_{\rm env}$ is located, especially
since during a common-envelope event the bottom of the outer convective
zone moves upwards in mass coordinate \citep{IvanovaThesis}. To help
characterize the instability of the envelope, we therefore also
introduce three derived quantities:

\begin{itemize} 

\item $m_{\rm uns}$ -- the mass coordinate above which
the envelope is unstable to ejection, such that $\langle \Gamma_1
(m_{\rm uns}) \rangle < 4/3$. 

\item $\Delta m_{\rm env,4/3}$ -- the mass
of the envelope where in each mass shell the local $\Gamma_1(m) < 4/3$.

\item $\Delta m_{\rm esc}$ -- the mass of the upper part of the envelope
in which the expansion velocities exceed the local escape velocity. 
Of course, the validity of hydrostatic stellar models {\it after} $\Delta m_{\rm esc}$
starts to increase is questionable. \end{itemize}

In this study we follow the evolution of the above quantities as we
inject thermal energy into the envelope of a stellar model.  Instead of
using time as the main independent variable we will often use the amount
of heat which has been added to the star by following both the total
heat energy added through the artificial heating, $E_{\rm heat}^{\rm
gross}$, and also the total net heat energy that the star received,
$E_{\rm heat}^{\rm net}$ -- this is the total heat energy added plus all
of the nuclear energy that was generated in the star during its heating,
$E_{\rm nuc}$,  less all the energy that was radiated away from the
surface of the star.  Another quantity that we trace is the change of
the total energy of the core, $\Delta E_{\rm core}$, in the part of the
star that is {\it below} $m_{\rm bot}$.

In a realistic CEE, the distribution of the heat input within the
envelope would not be a simple function of radius or mass, and the
details would probably depend on many initial conditions such as the mass
ratio of the two stars, the initial density profile in the envelope, the
degree of corotation at the onset of the in-spiral, how angular momentum
is transported through the common envelope and where this leads to
direct kinetic energy deposition, and more \citep[see, e.g.,
][]{mmh79,podsi01,Ric2012,Ivanova+2013Review}. One might therefore
explore a large parameter space in trying to evaluate the importance
of the heat distribution.  In this work we concentrate on two limiting
cases: (a) uniform specific heat input throughout the envelope
and (b) intense heating in a narrow mass range at the bottom of the
envelope.

Note that the outcomes of CEEs are typically estimated by using a standard
energy formalism which compares the available energy (in this case the
change in the orbital energy $\Delta E_{\rm orb}$) to the binding energy
of the envelope. The energy input is presumed to be utilised at some
efficiency called $\alpha_{\rm CE}\le 1$, although it has been common
for binary population synthesis codes to resort to ``efficiencies'' of
greater than unity (typically in lieu of an assumed additional energy
source). The broad physical picture underlying this formalism implicitly
assumes that the orbital energy is converted into kinetic energy of the
envelope. Whilst this is natural during a dynamical time-scale
plunge-in, conversion of orbital energy into kinetic energy during a
self-regulated spiral-in is definitely indirect.

If the internal energy of the envelope is included in the binding energy
calculation, then the standard energy formalism also assumes that the
internal energy of the envelope is converted into kinetic energy of the
outflow. Whether or not internal energy can be converted into kinetic
energy in this way is not established, even less whether the internal
energy would be converted with the same efficiency as the energy input
from orbital decay.

\section{Initial conditions and numerical code}

\begin{figure}
\includegraphics[width=90mm]{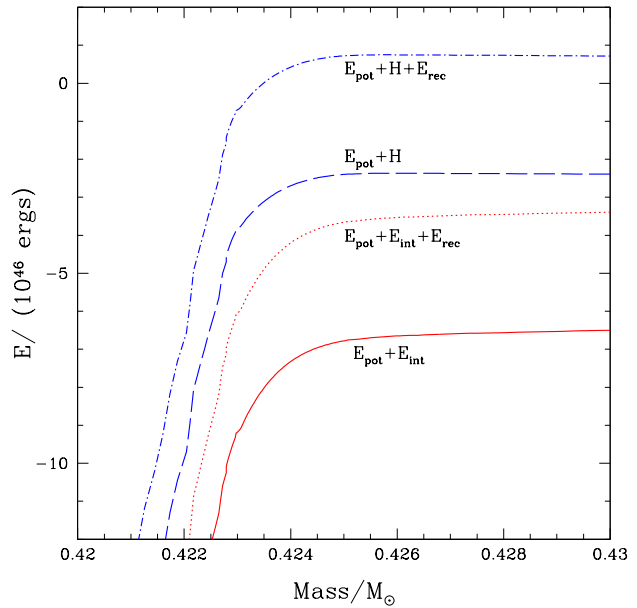} \\
\vskip-2.5cm
\caption{Energies as a function of $m_{\rm bot}$, in the 1.6 ${\rm{M}}_\odot$
giant with a radius of 100 $\rm {\rm{R}}_\odot$. 
$E_{\rm pot}$ -- the potential energy, $E_{\rm int}$ -- the internal energy,
$H$ -- enthalpy, $E_{\rm rec}$ -- the recombination energy reservoir.
For definitions see \S~\ref{sec:quant}.} 
\label{fig:star16_energy}
\end{figure}

For the main calculations in this paper we adopt the same initial
stellar model (we briefly describe tests on an alternative model in \S\ref{sec:common}). 
The model was chosen to be representative of the low-mass
giant stars which are commonly subject to CEEs.  This model is a red
giant of mass $M=1.6~{\rm{M}}_\odot$ and radius $R=100~{\rm{R}}_\odot$, with core mass
$M_{\rm core} = 0.422~{\rm{M}}_\odot$ (defined as the hydrogen-exhausted
region, where $X<10^{-10}$; the radius of the core is $\sim 0.02~{\rm R}_\odot$). 
The bottom of the convective envelope is at mass coordinate
$M_{\rm ce}=0.426~{\rm{M}}_\odot$ (the distance to the center is $\sim 0.8~{\rm{R}}_\odot$). 
The model was created by evolving a $M=1.6~{\rm{M}}_\odot$ zero-age
main-sequence star with metallicity $Z=0.02$ and hydrogen fraction
$Y=0.70$.

The hydrogen-burning shell in this star, as in all large-radius low-mass
giants, is very low in mass and vast in size. Since the radius
coordinate changes strongly with the mass coordinate, the potential and
thermal energies are strong functions of $m_{\rm bot}$. This can be seen
in Fig.~\ref{fig:star16_energy}, in which we show the energies in the
region of the star near in mass coordinate to the burning shell and to
the bottom of the convective zone. We note that the top curve in
Fig.~\ref{fig:star16_energy} becomes positive near the base of the
envelope; this suggests that if enthalpy defines the stability and
departure of the envelope, and if the recombination energy reservoir can
also be fully utilized, the envelope would have been unbound even before
any additional heating. Obviously, this joint condition requires that
the recombination energy must become available and hence something would
need to trigger recombination.

It is also important to realize that the total binding energy of the
star is about 250 times larger than the binding energy of the envelope.
Therefore any thermal feedback between the core and the envelope --
including changes in the energy output of the burning shell -- may
significantly alter the energy balance in the envelope: indeed, only a
1 per cent change in the core binding energy could potentially unbind the
envelope! Heating of the envelope during CEEs may well perturb the
interior layers sufficiently to cause such a feedback. {\it Hence it seems
unlikely that the changes in the energy of a stellar envelope during a
common-envelope event could be described properly by assuming that the
envelope is a closed system.}

The stellar models were evolved using the code and input physics
described in \cite{2004ApJ...601.1058I}. This code is capable of
performing both hydrostatic and hydrodynamic stellar evolution
calculations. However, for this study, our calculations do not employ
the ``dynamical term'' in the pressure equation (i.e.\ hydrostatic
equilibrium is assumed). Clearly this will alter our results, especially
smoothing over details of pulsational instabilities prior to ejection
\citep[see, e.g.,][]{1974ApJ...190..609W,Tuchman+1978,W+W1994}. 
However, we consider that the effect on the overall energy balance is
likely to be far smaller than our current uncertainty. For AGB envelope
ejection, \citet{W+W1994} find that including the dynamical terms leads
to pulsations and then envelope ejection marginally \emph{earlier} than
when assuming hydrostatic equilibrium (in which case ejection occurs
without pulsations), i.e. the dynamics of the ejection are substantially
different, but the occurrence of an instability is found for both
assumptions.  For this first study, we feel that consciously avoiding
pulsations should clarify the rest of the physics.

\section{Uniform heating of the envelope} \label{sec:uniform_heating}

\begin{figure} 
\vskip-1.2cm\includegraphics[width=90mm]{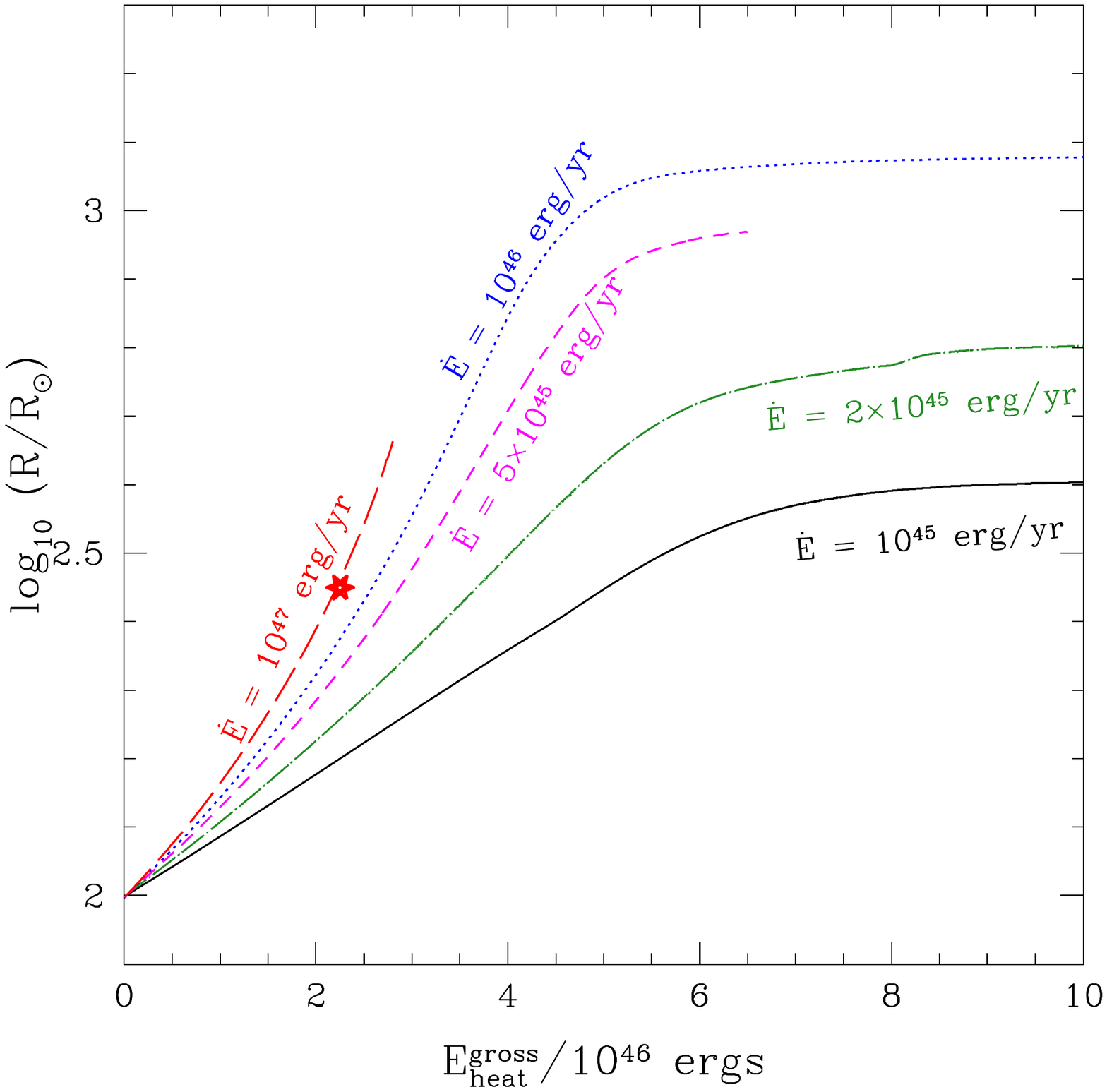} \\
\vskip-3.cm
\includegraphics[width=90mm]{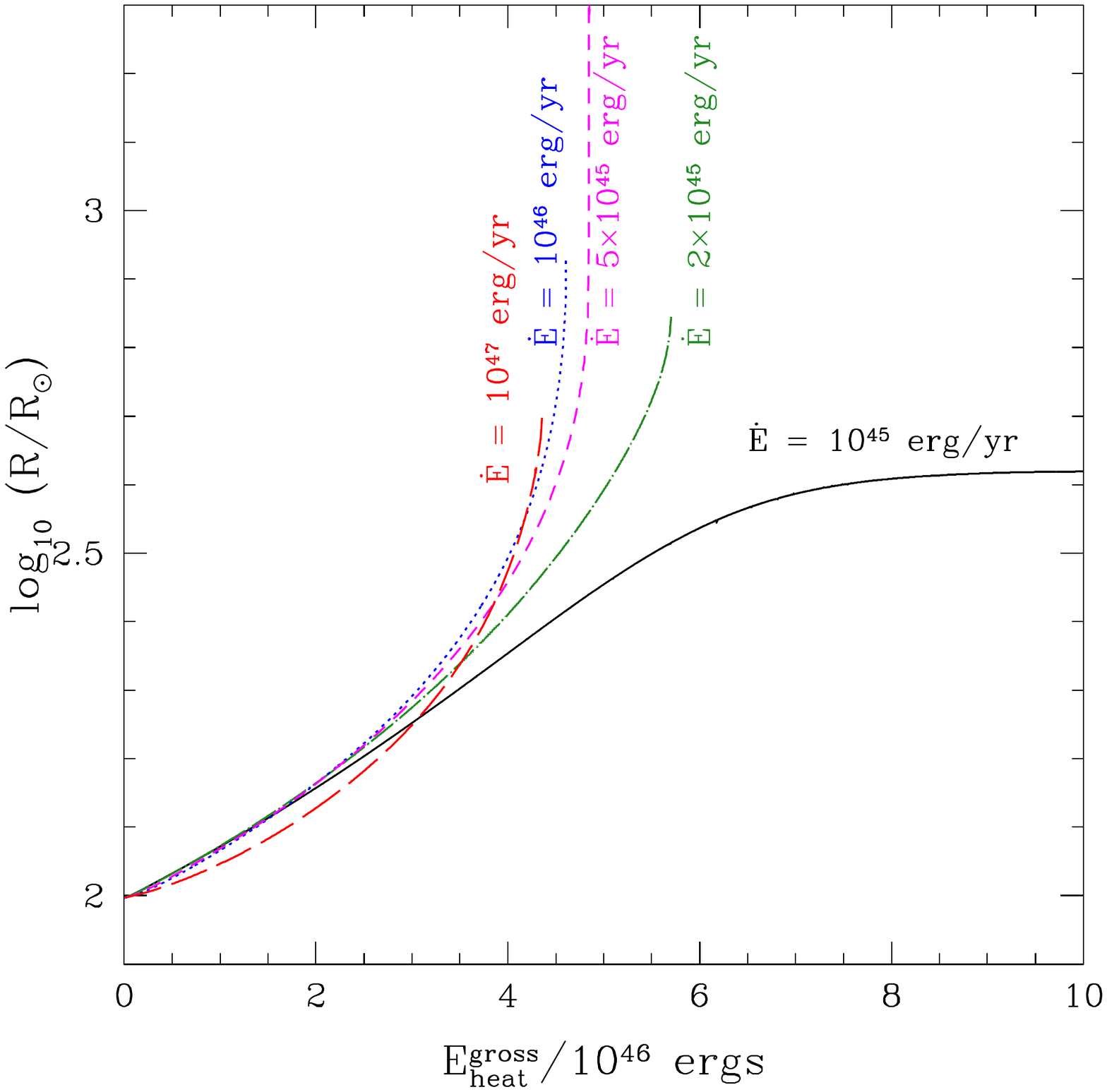} \\
\vskip-2.cm
\caption{The response of the star -- its expansion -- 
as a function of the amount of
heat injected into the envelope, for 5 heating rates. The top panel
shows the case when heat is evenly distributed by mass over the whole
envelope, and the bottom panel shows that case when the heat is
distributed evenly by mass into a shell with mass $0.1 {\rm{M}}_\odot$ at the
bottom of the initial convective envelope. For higher heating rates,
less integrated energy input is required before ejection occurs, whilst
for sufficiently low heating rates the stellar structure can adjust in
order to re-radiate all of the extra luminosity.} 
\label{fig:heat_con}
\end{figure}

\begin{figure} \includegraphics[width=90mm]{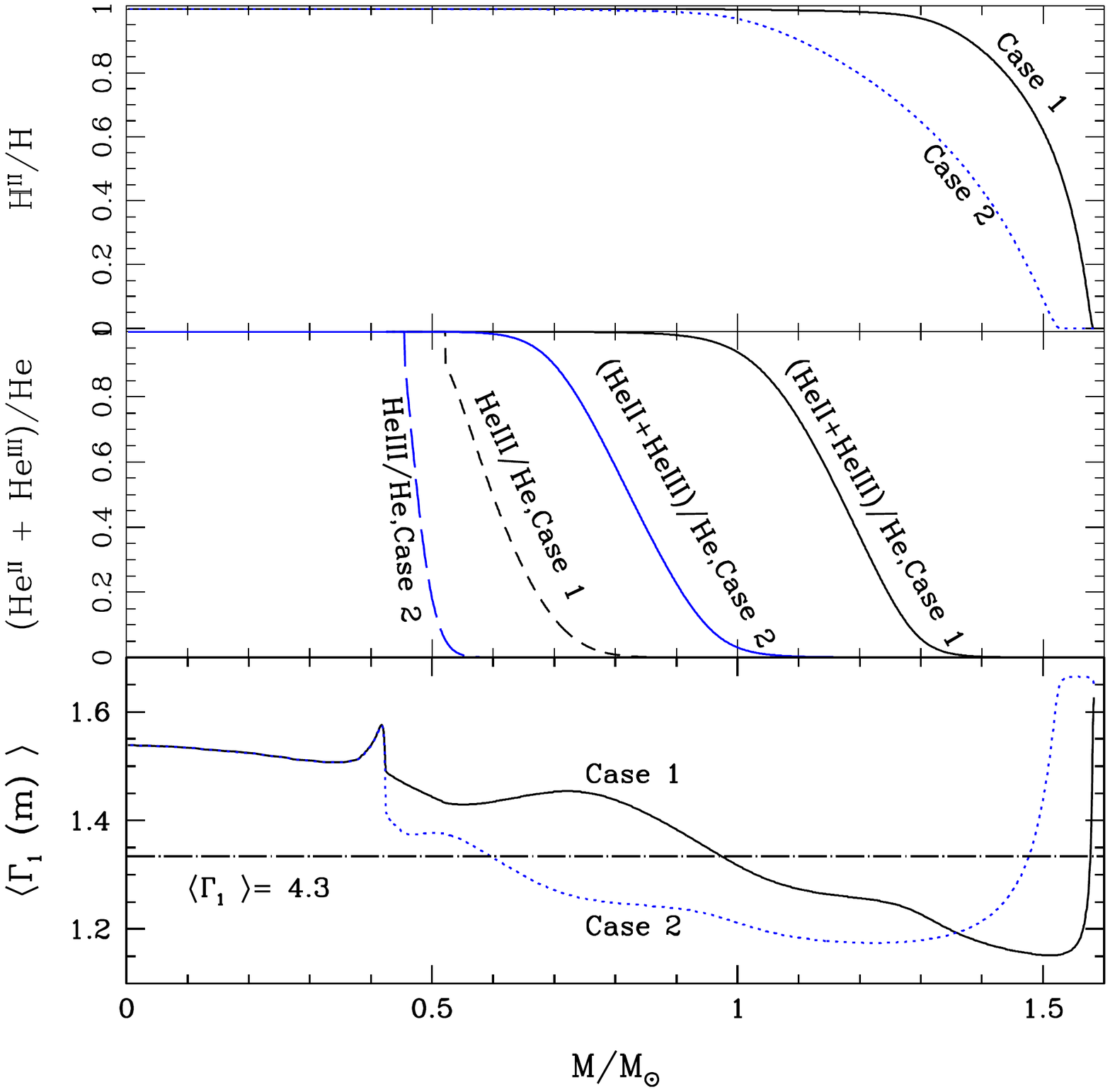} \\
\vskip-2.5cm
\caption{Envelope structure and partial ionisation of helium and hydrogen in Cases 1 and 2, uniform heating, at
steady-state, in both cases at $\Delta E_{\rm heat}^{\rm gross}=10\times10^{46}$ergs.
$\langle \Gamma_1 (m) \rangle $ is the pressure-weighted volume-averaged value of $\Gamma_1$
above the mass coordinate $m$.} 
\label{fig:case12_ion} 
\end{figure}

To study systematically the envelope response during a self-regulated
spiral-in, we first consider simple cases for which the heating is
uniformly distributed through the mass of the envelope (at various
constant rates). In this set of calculations we also adopt that no mass
is lost even if it has a velocity above the escape velocity. We introduced
the heating as an additional energy source, constant per gram, in the
entire initial convective envelope, with $m_{\rm bot}=M_{\rm ce}=0.426
{\rm{M}}_\odot$. The heating is not turned on and off sharply at the edges of
this region, but is smoothed such that the specific rate of additional
energy input decreases to zero over a transition zone with thickness
$0.01 {\rm{M}}_\odot$. We calculated sequences for five different rates of
heating $L_{\rm heat}$ ranging from $10^{45} {\rm erg\ yr}^{-1}$ to $10^{47} {\rm erg\ yr}^{-1}$
(see Table~\ref{tab:constant} and Fig.~\ref{fig:heat_con}).

We estimate that these rates of energy input cover a reasonable range of
values for a likely self-regulated spiral-in. This follows if we first
assume that this star is in a binary with a companion of $0.3 {\rm{M}}_\odot$,
and that the self-regulating spiral-in starts when the companion is
orbiting somewhat below the region where the original convective envelope was
located \citep[though during the spiral-in the envelope expands and so
the radiative layer might then be bigger than before the plunge, see
][]{IvanovaThesis,Han+2002}. In that case, the orbital energy at that
stage could be up to of order  $10^{48}$ ergs.  This CEE is
destined to eject the envelope leaving a binary behind; however, how
compact the final binary is  depends on when the envelope is ejected.
For a range of time-scales for a self-regulated spiral-in of between 10
and 1000 years, the heating luminosity can therefore be expected to be
between $\approx 10^{45}$ and $\approx 10^{47}$ erg per year. Clearly we
would not expect the heating rate to be constant in a real situation,
but to depend on the response of the envelope.
However, we do anticipate that in a realistic CEE at least the
frictional heating could be distributed throughout the
differentially-rotating envelope, and that the entire envelope could
be differentially rotating.

It is convenient to present this additional heat in two different units,
both in ergs per year (for ease of comparison with the binding energies
of the initial star) and in solar luminosities (to compare with the
unperturbed stellar luminosity).  The Eddington luminosity of our
initial model star is $L_{\rm Edd} = 5\times 10^{37}{\rm erg\ s}^{-1} \ 
\kappa_{\rm ph}^{-1} \ (M/{\rm{M}}_\odot)$, where $\kappa_{\rm ph}$ 
is the opacity of the photosphere in cm$^2$ g$^{-1}$. 
For comparisons of surface luminosities and
the energy input, we adopt Thompson scattering, not the actual material
opacity. So we write $L_{\rm Edd,TS}$. For our star, $L_{\rm Edd,TS}
\approx 64300 {\rm{L}}_\odot = 7.8\times 10^{45} {\rm erg\ yr}^{-1}$. Note that, in
two of our model sequences, the additional energy input to the star's
envelope appears to exceed $L_{\rm Edd,TS}$.   We also note that the
convective turnover time for the envelope of the unperturbed model was
calculated as approximately 495 days.

We now discuss the outcomes of our calculations, in order of increasing
rate of artificial heating.

\subsection{Cases 1 and 2: Readjustment and formal stability.}

\begin{center}
\begin{table*}
\begin{minipage}{180mm}
 \caption{Energies in the envelope: uniform heating.}
\label{tab:constant}
\fontsize{8}{9}\selectfont
{%
  \begin{tabular}{ *{4}{p{1.15cm}} *{5}{ p{0.65cm} } *{1}{ p{1cm} } *{3}{ p{0.8cm} } *{1}{ p{1cm} } }
\hline
{$\Delta E_{\rm heat}^{\rm gross}$}& 
{$\Delta E_{\rm heat}^{\rm net}$} & 
{$E_{\rm int}+E_{\rm pot}$} & 
{$H+E_{\rm pot}$}& 
{$E_{\rm rec}$} & 
{$E_{\rm rec}^{\rm H{\small II}}$} & 
{$E_{\rm rec}^{\rm He{\small II}}$} & 
{$E_{\rm rec}^{\rm He{\small III}}$} & 
{$E_{\rm kin}$} & 
{$\langle\Gamma_1(m_{\rm bot}) \rangle$} &
{$\Delta m_{\rm env, 4/3}$} &
{$m_{\rm unst}$} &
{$\Delta m_{\rm esc}$} &
{$L_*/{\rm L_\odot}$} \\
\hline
& \ \\
\multicolumn{14}{l}{Unperturbed star} \\
\ \\
 0.00& 0.00&-6.63&-2.35& 3.12& 2.05& 0.08& 0.99& 0.00& 1.62& 0.01& 1.56& 0.00&  1594. \\
\hline
& \ \\
\multicolumn{14}{l}{Case 1: $L_{\rm heat} = 10^{45} {\rm erg \ yr}^{-1} \equiv 8267 \times {\rm{L}}_\odot$} \\
& \ \\
%model heat_500_v3 (smooth over 0.01) 11930
1.99& 1.85&-4.65&-1.60& 3.00& 2.04& 0.12& 0.83& 0.00& 1.60& 0.04& 1.51& 0.00&  2835. \\
10.00& 4.35&-1.53&-0.50& 2.29& 1.90& 0.19& 0.20& 0.00& 1.49& 0.29& 0.97& 0.00&  9656. \\
\hline
& \ \\
\multicolumn{14}{l}{Case 2: $L_{\rm heat} = 2\times 10^{45} {\rm erg\ yr}^{-1} \equiv 16535 \times {\rm{L}}_\odot$} \\
& \ \\
%heat_250_v3 (0.01 smooth)  10000, 11000, 12005, 13005
 2.00& 1.91&-4.59&-1.59& 2.99& 2.04& 0.12& 0.83& 0.00& 1.60& 0.04& 1.49& 0.00&  3366. \\
 4.00& 3.46&-2.73&-0.96& 2.64& 1.99& 0.18& 0.47& 0.00& 1.54& 0.15& 1.26& 0.00&  7663. \\
 6.00& 4.26&-1.41&-0.49& 2.09& 1.78& 0.16& 0.15& 0.00& 1.47& 0.35& 0.84& 0.00& 14617. \\
 8.00& 4.51&-0.95&-0.32& 1.87& 1.66& 0.13& 0.07& 0.00& 1.43& 0.45& 0.65& 0.00& 16498.\\
10.00& 4.55&-0.83&-0.28& 1.77& 1.60& 0.12& 0.06& 0.00& 1.41& 0.47& 0.60& 0.00& 17626.\\
\hline
& \ \\
\multicolumn{14}{l}{Case 3 : $L_{\rm heat} = 5\times 10^{45} {\rm erg\ yr}^{-1} \equiv 41335 \times {\rm{L}}_\odot$} \\
& \ \\
%heat 100_v3 (0.01 smooth) 8400, 8800, 9001, 9101, 9111, 9121, 9150, 9171
 2.00& 1.95&-4.52&-1.58& 2.97& 2.04& 0.13& 0.80& 0.00& 1.59& 0.05& 1.49& 0.00&  4229. \\
 4.00& 3.58&-2.36&-0.85& 2.40& 1.90& 0.20& 0.30& 0.00& 1.55& 0.20& 1.13& 0.00& 18901. \\
 5.00& 3.91&-1.51&-0.55& 1.87& 1.59& 0.13& 0.14& 0.01& 1.51& 0.36& 0.71& 0.00& 38300. \\
 5.50& 3.94&-1.26&-0.46& 1.66& 1.44& 0.11& 0.11& 0.01& 1.49& 0.40& 0.61& 0.00& 40643. \\
 6.01& 3.97&-1.08&-0.39& 1.50& 1.33& 0.09& 0.09& 0.00& 1.48& 0.41& 0.56& 0.00& 40758. \\
 6.51& 3.99&-0.97&-0.35& 1.41& 1.26& 0.08& 0.07& 0.00& 1.47& 0.43& 0.53& 0.00& 40098. \\
\hline
& \ \\
\multicolumn{14}{l}{Case 4 : $L_{\rm heat} =  10^{46} {\rm erg\ yr}^{-1} \equiv 82672 \times {\rm{L}}_\odot = 1.3 \times {{L}}_{\rm Edd, TS}$ } \\
& \ \\
%heat 50 v3 8200, 8300, 8400, 8501, 8601, 8701, 8801
 2.00& 1.97&-4.48&-1.57& 2.96& 2.04& 0.14& 0.78& 0.00& 1.59& 0.04& 1.50& 0.00&  5028. \\
 3.00& 2.89&-3.35&-1.20& 2.73& 2.01& 0.19& 0.53& 0.00& 1.57& 0.09& 1.38& 0.00& 12115. \\
 4.00& 3.61&-2.21&-0.80& 2.29& 1.85& 0.18& 0.26& 0.05& 1.54& 0.26& 0.96& 0.00& 44039. \\
 5.00& 3.82&-1.37&-0.48& 1.64& 1.43& 0.08& 0.13& 0.04& 1.51& 0.35& 0.62& 0.00& 83600. \\
 6.00& 3.80&-1.00&-0.35& 1.25& 1.12& 0.05& 0.08& 0.01& 1.48& 0.42& 0.47& 0.00& 83391. \\
 8.00& 3.83&-0.70&-0.24& 0.99& 0.91& 0.03& 0.04& 0.00& 1.45& 0.42& 0.45& 0.00& 81374. \\
10.01& 3.89&-0.55&-0.19& 0.90& 0.85& 0.02& 0.03& 0.14& 1.42& 0.42& 0.44& 0.00& 82271. \\
\hline
& \ \\
\multicolumn{14}{l}{Case 5: $L_{\rm heat} =  10^{47} {\rm erg \ yr}^{-1} \equiv 826720 \times {\rm{L}}_\odot = 12.8 \times L_{\rm Edd, TS}$ } \\
& \ \\
% heat_5_v3 8100, 8150, 8200
2.00& 1.99&-4.45&-1.57& 2.96& 2.05& 0.15& 0.76& 0.10& 1.60& 0.02& 1.52& 0.00&  7363. \\
2.10& 2.10&-4.34&-1.54& 2.94& 2.05& 0.15& 0.74& 0.12& 1.59& 0.03& 1.51& 0.00&  8339. \\
2.25& 2.24&-4.17&-1.48& 2.91& 2.05& 0.16& 0.70& 0.19& 1.59& 0.04& 1.50& 0.00& 10099. \\
2.50& 2.49&-3.88&-1.38& 2.86& 2.04& 0.18& 0.64& 0.37& 1.58& 0.05& 1.45& 0.06& 15372. \\
2.65& 2.64&-3.70&-1.32& 2.82& 2.04& 0.18& 0.60& 0.61& 1.58& 0.07& 1.40& 0.09& 21110. \\
2.80& 2.78&-3.50&-1.24& 2.78& 2.04& 0.19& 0.55& 1.00& 1.57& 0.09& 1.35& 0.13& 31625. \\
\hline
\end{tabular}
}
\medskip
{\\ All energies in this Table are in units of $10^{46}$
    erg.  
$L_{\rm heat}$ is the rate of heating which was applied to the star.
$E_{\rm heat}^{\rm gross}$ records how much additional energy input
was provided to the star, whilst $E_{\rm heat}^{\rm net}$ 
is the resulting net energy gained by the star (accounting for the
nuclear energy input from the core and radiative losses from the
surface). 
$E_{\rm pot}$, $E_{\rm int}$ and $H$ are respectively the potential energy,
internal thermal energy and integrated enthalpy of the envelope.
The reservoir of recombination energy stored in the envelope at each
epoch is given by $E_{\rm rec}$, with the contributions from
ionised hydrogen, singly-ionised Helium and doubly-ionised Helium
correspondingly given as $E_{\rm rec}^{\rm H{\small II}}$, 
$E_{\rm rec}^{\rm He{\small II}}$ and  $E_{\rm rec}^{\rm He{\small III}}$.
$E_{\rm kin}$ is the kinetic energy of the envelope.
$m_{\rm bot}$ is the mass coordinate of the base of the envelope. 
$\langle \Gamma_1 (m_{\rm bot}) \rangle $ is the pressure-weighted volume-averaged value of $\Gamma_1$
in the envelope.  $\Delta m_{\rm env,4/3}$ is the mass
of the envelope where in each mass shell $\Gamma_1(m) < 4/3$ (locally).
$\Delta m_{\rm esc}$ is the mass of the upper part of the envelope
in which the expansion velocities exceed the local escape velocity.
$L_*$ is the surface luminosity.
For definitions of these quantities see also \S~\ref{sec:quant}. 
}
\end{minipage}
\end{table*}

\end{center}

\emph{Case 1.} This model star was heated at a rate of $10^{45}$ ergs
per year. The stellar structure expanded until the star reached the
surface luminosity at which it radiates the same amount of energy as the
artificial heating source and the burning hydrogen shell provide
together. Most of the initially-doubly-ionised helium has recombined in
the envelope above $m_{\rm bot}$, while only several  per cent of the
initially-ionised hydrogen has done so (see Table~\ref{tab:constant}).
Although in Table~\ref{tab:constant} we show only the model with $\Delta
E_{\rm heat}^{\rm gross}=10\times 10^{46}$ ergs, the star kept evolving
in anunchanged state until the total artificial energy input had been at
least $\Delta E_{\rm heat}^{\rm gross}=30\times 10^{46}$ ergs, i.e.
about 5 times the initial binding energy of the star. At that point we
stopped the simulations.

However, except for a tiny mass of about $0.03 {\rm{M}}_\odot$ near to the
surface, in which the local $\Gamma_1> 4/3$, a significant fraction of
the envelope is mechanically unstable (see the Table~\ref{tab:constant}
and Figure~\ref{fig:case12_ion}) and any perturbation may drive its
ejection. Since the orbiting binary will very probably cause such a
perturbation, it seems reasonable to expect that this part of the
envelope could be ejected instead of settling into an eternal
self-regulating spiral-in.

\emph{Case 2.} As in Case 1, the star approaches a stable state in
which it radiates away the combined nuclear and heating luminosity.
For this higher heating rate, the mass of the envelope that is
potentially dynamically unstable is bigger, but so is the mass of the
near-surface region with local $\Gamma_1> 4/3$ (see Figure~
\ref{fig:case12_ion}).  This stable near-surface region is more
massive because the hydrogen partial ionisation zone, \ionh, has moved
inwards. In Figure~\ref{fig:case12_ion} we show the location of the
partially ionised layers and the way that $\langle \Gamma (m) \rangle
$ changes in the envelope. This suggests that any potential dynamical
instability is driven by a low $\Gamma_1$ in the zone of partial
ionisation of hydrogen and the first partial ionisation zone of
helium, He{\small{II}}. Note that in Case 2 \emph{almost the entire
  envelope} has its helium incompletely ionised, i.e.\ the helium
partial ionisation zone is at the bottom of the envelope.  This is
also traced by the change in $E_{\rm rec}$ -- almost all of the
recombination energy initially stored in doubly ionised helium,
He{\small{III}}, has been released. Another interesting quantity is
$\Delta m_{\rm env, 4/3}$ -- this is how much of the envelope mass has
its local $\Gamma_1 < 4/3$ -- which appears to trace the thickness of
the partially ionised hydrogen layer plus the mass where
He{\small{II}} $>0.01$. Almost the entire envelope (all the mass above
$0.6~{\rm{M}}_\odot$) is dynamically unstable.

\subsection{Case 3: Readjustment and then instability}

These calculations become unstable after the last model shown in
Table~\ref{tab:constant}. In part this instability may well be
numerical, and we cannot be sure that it is not entirely numerical. The
calculations seem to become undecided over whether the models should
converge to either a smaller or a larger radius, each with differently
distributed ionisation zones. This numerical instability can be
suppressed for some time by fine-tuning of the time-steps. However, we
consider that there is a physical reason leading to this instability.
The initial dynamical time-scale for the star, before the artificial
heating, is $\tau_{\rm dyn}=0.04$ year. As the star expands in
response to the energy input,  $\tau_{\rm dyn}$ also increases. For
example, the steady-state expanded stars in Cases 1 and 2 have
$\tau_{\rm dyn}\approx0.5$ year; in Case 3, at the plateau state,
$\tau_{\rm dyn}\approx1.1$ year. Each gram of the material in this
stellar envelope is heated by $\sim 2 \times 10^{12}$ erg per gram per
year, while the local specific binding energy of the envelope material
at this moment is only $\sim 3\times 10^{12}$ ergs. So in roughly one
dynamical time the outer layers are being heated by more than their
binding energy, i.e.\ the heating of the outermost layers has become
dynamical. In addition, most of the envelope mass is already dynamically
unstable. We feel this combination indicates physical envelope ejection
is likely, not just a numerical instability. Nonetheless, confirming
  that this instability is physical will require future
  calculations in which we follow the dynamics of the envelope.

\begin{figure*}
\includegraphics[width=88mm]{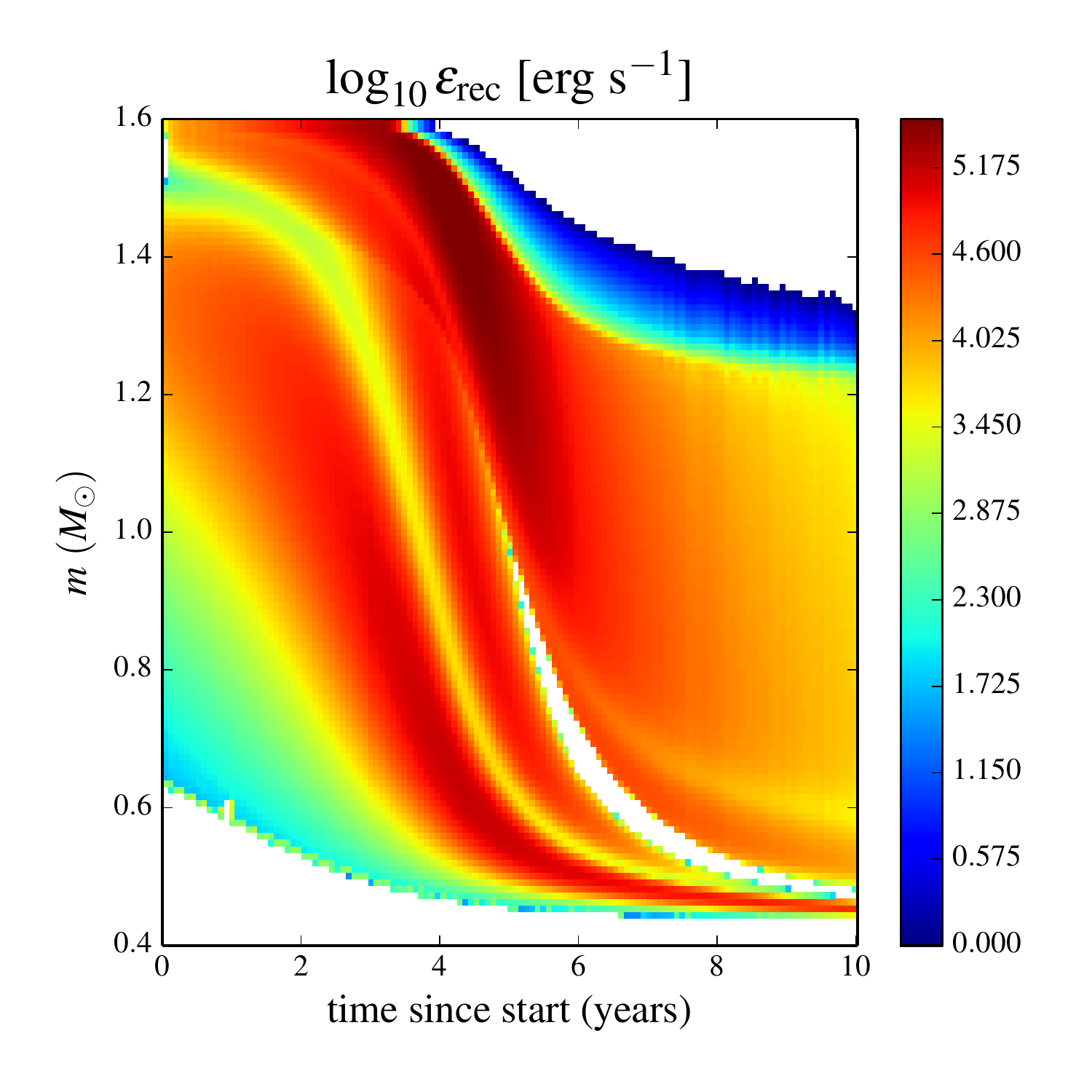}
\includegraphics[width=88mm]{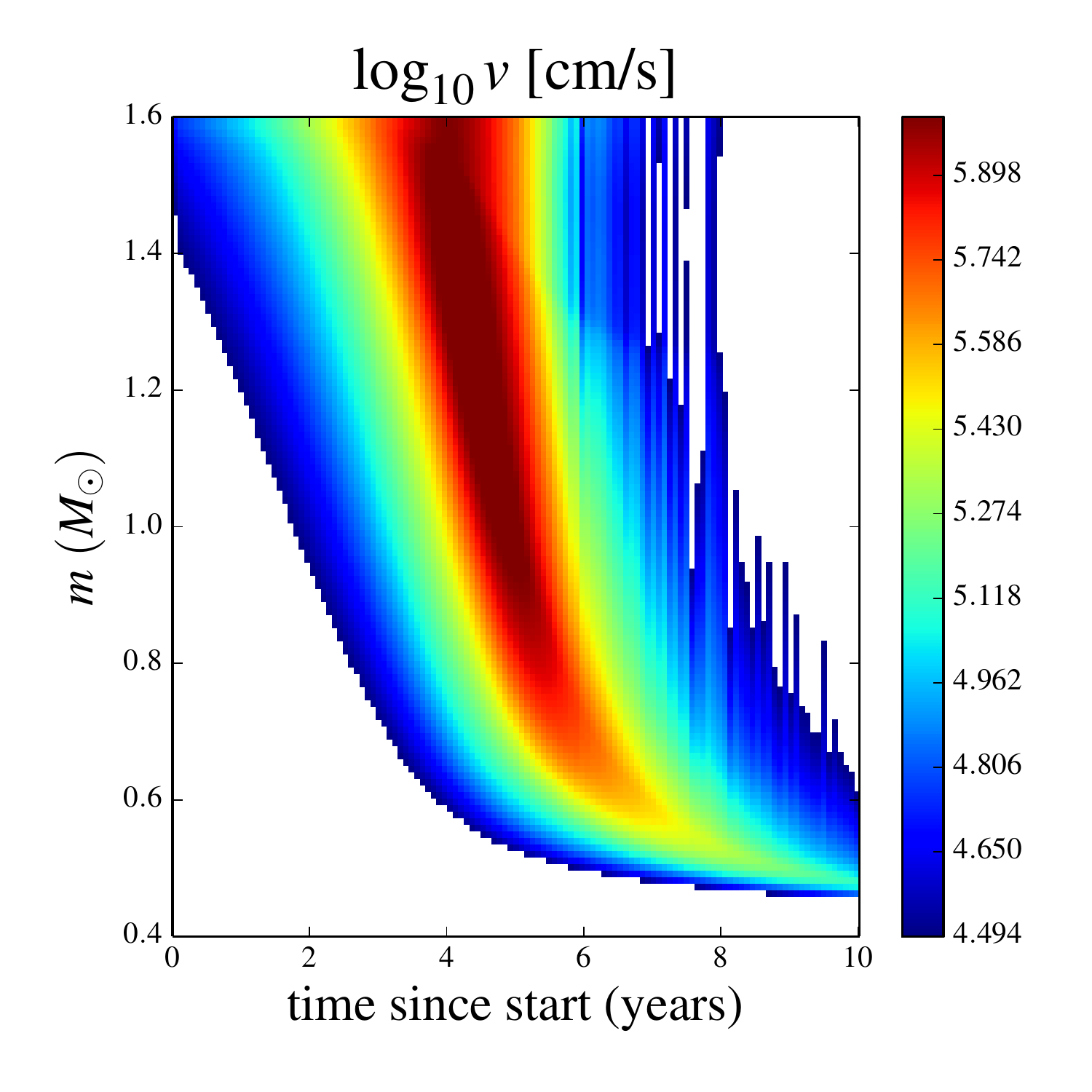}
\caption{The rate of recombination energy release (the left panel) and
expansion velocities (the right panel) in the Case 4 uniform heating. }
\label{fig:2d_erec_case4_uniform} 
\end{figure*}

\subsection{Case 4: A Super-Eddington star}

The heating rate for this sequence exceeds $L_{\rm Edd, TS}$, but this
star is able to expand  to a luminosity $L>L_{\rm Edd, TS}$ (see
Fig.~\ref{fig:heat_con} and Table\ref{tab:constant}). This has become
possible because the opacities at the surface of this very cold star
($T_{\rm eff}<3000K$) are {\it lower} than for Thompson scattering.

In this example, recombination drives the expansion of the star.
Table\ref{tab:constant} shows that the total release of recombination
energy becomes comparable to the rate of heating (e.g.\ the
recombination energy release is 65 per cent of the total heating luminosity -- 
$0.65\times10^{46}$ erg yr$^{-1}$ -- between the moments when
$E_{\rm heat}^{\rm gross}$ increases from 4 to $5\times 10^{45} \rm erg$). That
recombination energy is released in a smaller mass than the heating
luminosity and so is locally dominant. We note that this energy 
mostly comes from hydrogen recombination and is dominant in the mass
regions which expand at the fastest rate (see
Figure~\ref{fig:2d_erec_case4_uniform}\footnote{Plots
which present 15 different internal quantities for 13 models can be found in
the Supplementary Online Material. Five calculations at different heating rates are shown for each of
uniform and bottom heating, as well as models X6, X8 and MS015. The
plots show luminosity, entropy, the rate of the
recombination energy release, the distribution of convective and radiative
zones, opacity, the radial coordinate of each mass point, velocity, the
ratio of the local luminosity to the local Eddington luminosity in 
convective regions in which convection would need to be supersonic to carry the predicted convective flux,
the ``work'' term
in the gravitational energy $P/\rho^2 d\rho/dt$, the time derivative of the
internal energy $dU/dt$, the local value of $\Gamma_1$, the 
pressure-weighted volume-averaged $\langle \Gamma_{m} \rangle$ integrated inwards from the surface, and
the ionisation fractions of hydrogen, of singly ionised helium and of doubly
ionised helium.}).

This model is on the edge of instability; almost the entire envelope above
$m_{\rm bot}$ is unstable, and the star becomes unstable soon after the
last model shown in the Table. The fact that Case 4 appears to be more
stable than Case 3 is in part numerical -- due to better fine-tuning of
the time-steps -- and in part because a larger fraction of the outer
envelope is radiative: this star has a radiative envelope above $1.44
{\rm{M}}_\odot$, while in Case 3 the outer radiative envelope starts at $1.51
{\rm{M}}_\odot$.  So it may be that for uniform heating of the envelope,
instability and ejection is not a monotonic function of the rate of
energy input -- because of the way the heating of the outer regions
alters the structure of the outer envelope.

\subsection{Case 5. Dynamical heating.}

For our unperturbed star, one needs to add  $\sim 3\times 10^{13}$ ergs
per gram to the material in the outer envelope to unbind that matter.
The uniform heating rate of $10^{47}$ erg yr$^{-1}$ corresponds to
heating of the envelope by $4.3 \times 10^{13}$ erg per gram per year.
It is therefore not surprising that this heating will result in the 
dynamical ejection of the surface layers when the star has expanded such
that the dynamical time-scale approaches a year.

We can write the star's expansion with time as a function of heating
energy, using the fact that in this case of rapid heating $\Delta E_{\rm
heat}^{\rm net}/dt\approx \Delta E_{\rm heat}^{\rm gross}$ (which we
have confirmed using the calculations). In the same units as in
Fig.~\ref{fig:heat_con}, we have:

\begin{equation} \frac{d R}{dt} = \frac{d R/{\rm{R}}_\odot}{d E}  \frac{d
E}{\rm yr} {{\rm{R}}_\odot\ \rm  yr^{-1}} =    10 \times \frac{d
R/{\rm{R}}_\odot}{\Delta E_{\rm heat}^{\rm gross}/10^{46} {\rm erg}} {\rm{R}}_\odot\
{\rm yr^{-1}} \ .\end{equation}

The surface escape velocity for our star is $v_{\rm esc} = 7.8\times
10^7 \times ({R/{\rm{R}}_\odot})^{-1/2} {\rm cm\ s}^{-1} $. In the unperturbed star
$v_{\rm esc} =78\,$km\ s$^{-1}$. As the star expands, the surface escape velocity
decreases, and can be written as:

\begin{equation} v_{\rm esc} = \frac{35,300}{\sqrt{R/{\rm{R}}_\odot}}
R_{\odot}{\rm yr^{-1} } . \end{equation}

We can thereby determine when the star's expansion is faster than its surface
escape velocity:

\begin{equation} \sqrt{\frac{R}{R_{\odot}}} d (\frac{R}{R_{\odot}}) >
3530 \frac{dE_{\rm heat}^{\rm gross}}{10^{46}~{\rm erg}} \ .\label{eq_esc}
\end{equation}

From this we can estimate that free streaming
should be expected to start before the star has expanded to $\sim 900\,
{\rm{R}}_\odot$, as the initial binding energy of the envelope is less than $7\times
10^{46}$ erg. This estimate is supported by our calculations, as shown
in Fig.~\ref{fig:heat_con}, in which the star symbol marks where  the radius
derivative satisfies Eq.~\ref{eq_esc} above. This moment, at which the
star's surface layers start to expand at a speed comparable to the
star's current escape velocity, occurs well before the radius reaches
900 ${\rm{R}}_\odot$. At later times, deeper layers reach escape
velocity. By the last model shown in Fig.~\ref{fig:heat_con}, $\sim 0.13
{\rm{M}}_\odot$ of the envelope had a velocity higher than the local escape
velocity.

We also note that the final envelope is less recombined than in Cases 1
and 2, i.e., more energy is still stored in the ionised plasma when the
dynamical instability begins.

\subsection{Consequences for CEE from uniform heating}

\begin{itemize} 
\item In none of the models was the total \emph{net}
heat added to the envelope greater than the initial binding energy of
the envelope. 
\item The additional heat input also leads to a change in
the binding energy of the interior.  There is no simple but accurate
energy balance that considers only the envelope, and the energy balance
would strongly depend on the time-scale of the self-regulated spiral-in.
\item The outcome depends on the \emph{rate} at which heat was provided,
not on the total energy added. Faster heating causes the stellar
envelope to begin streaming away at lower $\Delta E_{\rm heat}^{\rm
net}$. 
\item For constant energy deposition rate then, if the heating
luminosity is significantly lower than star's Eddington luminosity, the
star will adjust to radiate away all of the additional energy input.
\item  The star's envelope can recombine when it attempts to reach a
``steady'' state, i.e.\ readjusting to try to re-radiate the additional
energy input. Then the helium will have recombined through most of the
envelope. Since the second partial ionisation zone of helium is rather
thick in mass, this seems likely to lead to Cepheid-type
pulsations; note that we can not obtain normal Cepheid pulsations naturally
with the hydrostatic code we used. \end{itemize}

One large inconsistency with this model is that the heating is uniform
all the way to the surface. This causes heating of the outer envelope on
a dynamical time-scale, which leads to instabilities. In a more realistic
situation, this surface heating would not occur.

\begin{center}
\begin{table*}
\begin{minipage}{180mm}
 \caption{Energies in the envelope: bottom heating.}
\label{tab:constant_bot}
\fontsize{8}{9}\selectfont
{%
  \begin{tabular}{ *{4}{p{1.15cm}} *{5}{ p{0.65cm} } *{1}{ p{1cm} } *{3}{ p{0.8cm} } *{1}{ p{1cm} } }
\hline
{$\Delta E_{\rm heat}^{\rm gross}$}& 
{$\Delta E_{\rm heat}^{\rm net}$} & 
{$E_{\rm int}+E_{\rm pot}$} & 
{$H+E_{\rm pot}$}& 
{$E_{\rm rec}$} & 
{$E_{\rm rec}^{\rm H{\small II}}$} & 
{$E_{\rm rec}^{\rm He{\small II}}$} & 
{$E_{\rm rec}^{\rm He{\small III}}$} & 
{$E_{\rm kin}$} & 
{$\langle\Gamma_1(m_{\rm bot}) \rangle$} &
{$\Delta m_{\rm env, 4/3}$} &
{$m_{\rm unst}$} &
{$\Delta m_{\rm esc}$} &
{$L_*/{\rm L_\odot}$} \\
\hline
& \ \\
\multicolumn{14}{l}{Unperturbed star} \\
\ \\
0.00& 0.00&-6.63&-2.35& 3.12& 2.05& 0.08& 0.99& 0.00& 1.62& 0.01& 1.56& 0.00&  1594. \\
\hline
& \ \\
\multicolumn{14}{l}{Case 1: $L_{\rm heat} = 10^{45} {\rm erg\ yr}^{-1} \equiv 8267 \times {\rm{L}}_\odot$} \\
& \ \\
%p_5_v3 10010, 12020, 15342
 2.00& 1.87&-4.63&-1.60& 2.98& 2.04& 0.12& 0.81& 0.00& 1.60& 0.04& 1.51& 0.00&  2625. \\
 4.00& 3.36&-2.93&-1.01& 2.73& 2.02& 0.18& 0.53& 0.00& 1.55& 0.11& 1.36& 0.00&  4691. \\
10.00& 4.40&-1.39&-0.47& 2.20& 1.86& 0.18& 0.17& 0.00& 1.47& 0.33& 0.92& 0.00&  9625. \\
\hline
& \ \\
\multicolumn{14}{l}{Case 2: $L_{\rm heat} = 2\times 10^{45} {\rm erg\ yr}^{-1} \equiv 16535 \times {\rm{L}}_\odot$} \\
& \ \\
% p_250_v2 9005, 10010, 10512, 10682, 10732
 2.00& 1.95&-4.54&-1.57& 2.98& 2.04& 0.13& 0.80& 0.00& 1.60& 0.04& 1.50& 0.00&  2673. \\
 4.00& 3.66&-2.53&-0.88& 2.63& 2.00& 0.19& 0.44& 0.00& 1.54& 0.15& 1.29& 0.00&  5540. \\
 5.00& 4.31&-1.55&-0.53& 2.28& 1.90& 0.19& 0.20& 0.00& 1.49& 0.31& 1.00& 0.00&  9383. \\
 5.50& 4.51&-1.05&-0.36& 1.97& 1.73& 0.14& 0.10& 0.00& 1.44& 0.44& 0.71& 0.00& 13929. \\
 5.70& 4.53&-0.77&-0.26& 1.70& 1.55& 0.09& 0.05& 0.04& 1.39& 0.50& 0.52& 0.00& 22661. \\
\hline
& \ \\
\multicolumn{14}{l}{Case 3 : $L_{\rm heat} = 5\times 10^{45} {\rm erg\ yr}^{-1} \equiv 41335 \times {\rm{L}}_\odot$} \\
& \ \\
% model p_100, 8430, 8940, 9150, 9190, 9230, 9270, 9290
 2.00& 1.98&-4.52&-1.58& 2.96& 2.04& 0.13& 0.79& 0.00& 1.59& 0.04& 1.50& 0.00&  2655. \\
 3.99& 3.85&-2.29&-0.80& 2.56& 1.99& 0.20& 0.37& 0.00& 1.53& 0.18& 1.23& 0.00&  6328. \\
 4.51& 4.29&-1.59&-0.55& 2.29& 1.91& 0.18& 0.20& 0.01& 1.49& 0.31& 1.00& 0.00& 10496. \\
 4.61& 4.37&-1.44&-0.50& 2.21& 1.87& 0.17& 0.16& 0.01& 1.48& 0.34& 0.89& 0.00& 12324. \\
 4.71& 4.44&-1.27&-0.44& 2.11& 1.82& 0.16& 0.13& 0.02& 1.46& 0.41& 0.80& 0.00& 15208. \\
 4.81& 4.50&-1.06&-0.37& 1.96& 1.74& 0.13& 0.09& 0.10& 1.44& 0.47& 0.68& 0.00& 21987. \\
 4.86& 4.52&-0.91&-0.31& 1.81& 1.65& 0.10& 0.07& 0.71& 1.41& 0.50& 0.54& 0.29& 33605. \\
\hline
& \ \\
\multicolumn{14}{l}{Case 4 : $L_{\rm heat} =  10^{46} {\rm erg\ yr}^{-1} \equiv 82672 \times {\rm{L}}_\odot = 1.3 \times {{L}}_{\rm Edd, TS}$ } \\
& \ \\
% p_50 8350, 8830, 9150, 9250, 9270, 9290
 2.01& 2.00&-4.48&-1.57& 2.96& 2.04& 0.13& 0.78& 0.00& 1.59& 0.04& 1.50& 0.00&  2637. \\
 4.01& 3.94&-2.11&-0.75& 2.49& 1.98& 0.20& 0.31& 0.01& 1.52& 0.20& 1.16& 0.00&  7268. \\
 4.50& 4.38&-1.33&-0.47& 2.14& 1.85& 0.16& 0.14& 0.10& 1.47& 0.41& 0.80& 0.00& 17473. \\
 4.60& 4.45&-1.10&-0.39& 1.98& 1.76& 0.12& 0.09& 0.50& 1.44& 0.49& 0.66& 0.14& 28370. \\
 4.62& 4.46&-1.04&-0.37& 1.93& 1.73& 0.11& 0.08& 1.32& 1.43& 0.49& 0.61& 0.37& 33629. \\
 4.64& 4.47&-0.96&-0.34& 1.85& 1.69& 0.10& 0.07& 3.21& 1.41& 0.51& 0.54& 0.63& 42871. \\
\hline
\end{tabular}
}
\medskip
{\\ All energies in this Table are in units of $10^{46}$
    erg.  For descriptions of the variables see Table \ref{tab:constant} and
  \S~\ref{sec:quant}. }
\end{minipage}
\end{table*}

\end{center}

\section{Bottom heating} \label{sec:bottom_heating}

For this set of simulations, the additional energy input was introduced
into the bottom $0.1 {\rm{M}}_\odot$ of the initial convective envelope.  This
situation more closely resembles the local heating during a phase of
self-regulating spiral-in, albeit it lacks subsidiary heating
of the surface layers that would be present in a more realistic case of
CEEs.  We injected the same total amount of energy as in the case of
uniform envelope heating, again uniformly distributed in mass but only
spread over the $0.1 {\rm{M}}_\odot$ shell.

Overall, this more concentrated bottom heating is significantly more
effective in causing the stellar envelope to stream out at a smaller
imposed $\Delta E_{\rm heat}^{\rm gross}$.

The upper and lower extreme heating rates (Case 1 and Case 5) produce
outcomes which are qualitatively similar to the corresponding
simulations with uniform heating. In Case 1 the star adjusts to enable
eternal self-regulation, and in Case 5 the envelope is dynamically
ejected. All the intermediate simulations (Cases 2, 3 and 4) also lead
to dynamical ejection after $\Delta E_{\rm heat}^{\rm net}\approx
4.5\times10^{46}$.  At that moment, $\Delta m_{\rm env, 4/3}\approx 0.5$
for all models. The envelopes have $\Gamma_1<4/3$ down to 0.5 ${\rm{M}}_\odot$
and $\Gamma_1=1.4$ down to $m_{\rm bot}$.

The development of the envelope expansion is smooth and does not cause
obvious numerical problems until the local velocities exceed their local
escape velocities; at this point, we can not fully trust the stellar
models anymore  (although we still list the output in the table). We
also note that the rate of the total energy release provided by
recombinations at this moment exceeds $L_{\rm Edd, TS}$ (see
Fig.~\ref{fig:rec_lum} and also the discussion in \S\ref{sec_rec}). 
Strictly speaking, the models may become unreliable somewhat earlier
than that, when the local expansion velocities exceed the convective
velocities.

\begin{figure}
\includegraphics[width=85mm]{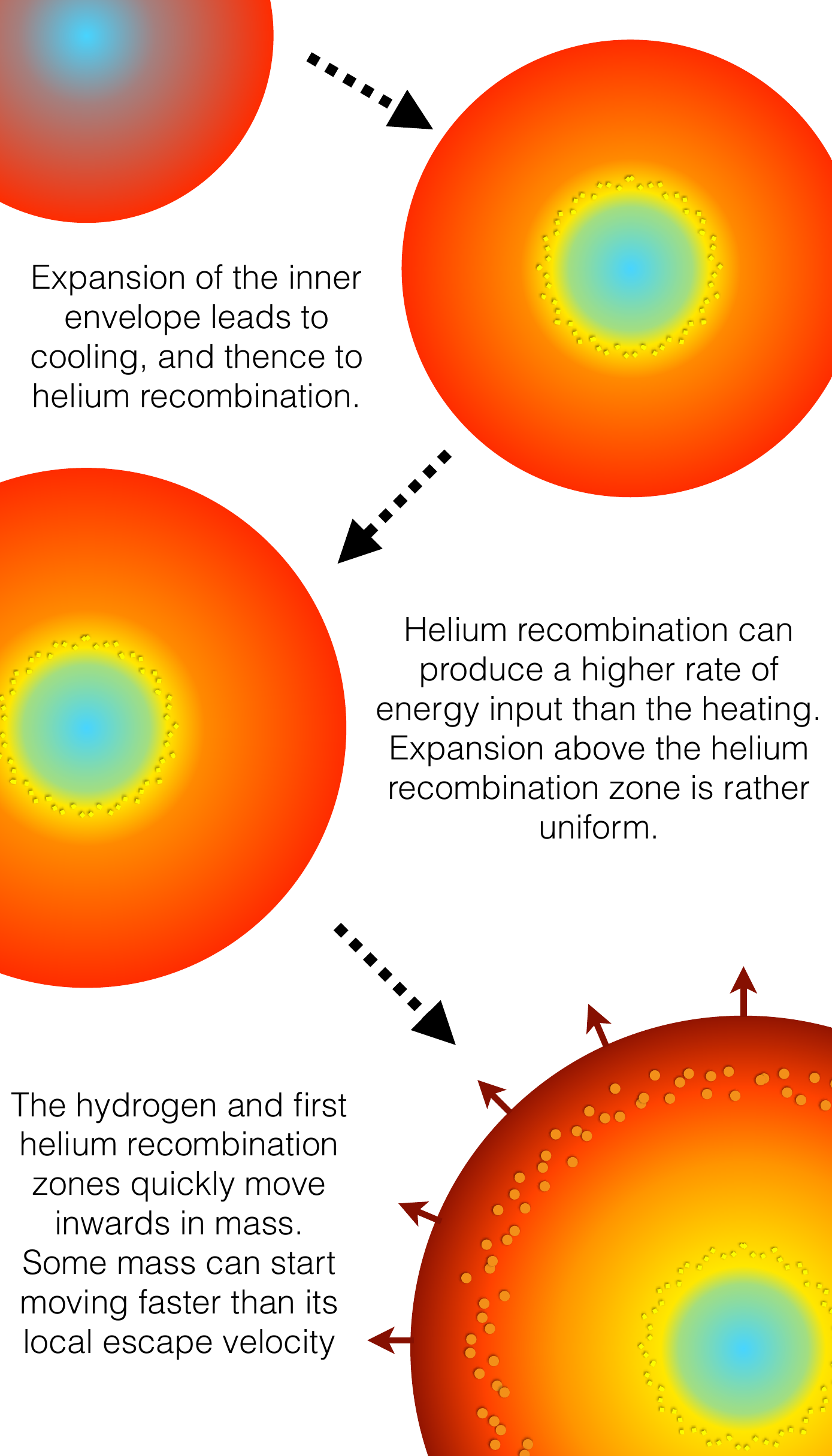} 
\caption{The stages of envelope ejection in the bottom-heating
case.} 
\label{fig:schematic} 
\end{figure}

We can identify the following stages of envelope ejection
(see also Figure~\ref{fig:schematic}): 

\begin{itemize}
\item Expansion of the inner envelope leads to cooling. 
When the cooling is sufficient, this causes helium
recombination.

\item Helium recombination proceeds and can sometimes produce a higher
rate of energy input than the heating which led to the recombination:
$\dot E_{\rm rec} > \dot E_{\rm heat}^{\rm gross}$. Heating of the layers where helium
has already recombined causes more rapid expansion than before recombination.  
The envelope above the helium recombination zone expands rather uniformly.

\item Hydrogen recombination then moves inwards in mass as the envelope expands. 
This can also release energy at a higher rate than the
heating. Because this occurs in the outer parts of the
  envelope, they are the most affected by this release of energy, and
  so the outer layers begin to expand rapidly.
Hydrogen recombination and first helium recombination zones are quickly moving inwards
in mass. This is the stage during which the envelope acquires a kinetic
energy above $10^{44}$ erg and some mass can start moving with a speed
above its local escape velocity. 
\end{itemize}

For all of the models calculated for this paper, almost all of the
energy which is initially stored in ionised helium was used to help
expand the envelope. We are therefore tempted to conclude that it is
normally true that the vast majority of helium recombination energy is
useful in envelope expansion.  However, a smaller fraction of the
energy stored in ionised hydrogen is used to expand the envelope -- we
estimate between a few per cent and 60 per cent -- and we stress that
hydrogen recombination is \emph{less} efficient in the case of runaway
envelopes than in cases of self-regulated expansion (see also the
discussion on the role of recombination in \S\ref{sec_rec}).

We find that in all models which ended with a runaway, the
rate of  energy input from the recombination of
helium at each epoch exceeds that from hydrogen recombination. When
the recombination zone approaches the bottom-heated layer then the
relative rate of recombination energy release between models follows
the relative differences in the heating rate between those models. 
For the self-regulated models, the maximum rate of
recombination energy release from helium also changes with the heating
rate, albeit for those models the rate of energy release from hydrogen
recombination 
can be higher than that from helium at the same instant.
This leads us to the conclusion that one of the most important effects
of the heating is to trigger helium recombination and that the local rate of
that helium recombination depends strongly on the heating rate.

\subsection{Testing for the moment of instability}

Since the instability takes a finite time to develop, we tested whether
our heating had been applied for longer than necessary to trigger the
instability.  Perhaps the constant heating in our earlier simulations
had continued even after the envelope had become unstable?

One could expect a star to develop dynamically instability on its
dynamical time, which is about a year for our expanded stars. For Case
4, a year of heating implies a different gross energy input by $\approx
10^{46}$ erg!

For this test we took several stars from the heating sequence and let
them evolve freely, without further heating, and studied whether the
star continued to expand and eject the envelope or remained bound and
contracted back towards the initial configuration.  In the Case 4
bottom-heating sequence, the star with $\Delta E_{\rm gross}^{\rm
  heat}=4.62$ (see Table~\ref{tab:constant_bot}) is the last star that
contracts when heating is switched off. Later models in the sequence
keep expanding. This demonstrates that, in this case, the moment when
the star has become unstable and our final model are not very
different.

\subsection{Testing for the role of the recombination energy input}

To investigate the role of the recombination energy release in the
outcome of heating, we tested what happens in an imaginary situation
in which recombination energy cannot be used. We re-calculated Cases
2,3 and 4 with bottom heating, but instantaneously removed the
recombination energy from the gravitational energy source in the
stellar structure equations.  However, we anticipate that removal of
the recombination energy could introduce numerical instabilities;
therefore these calculations should be considered less trustworthy
than our main results.

All of these modified calculations produced results which were
different from the unmodified case.  For the bottom-heated Case 4 with
no recombination energy input -- where helium recombination would have
played a smaller role in the total energy budget than in Case 2 -- we
found slower radius expansion for equivalent $\Delta E_{\rm heat}$
than for the Case 4 calculations which included recombination energy
input (for either uniformly-distributed or bottom-concentrated
heating). However, the expansion of this star is still very fast
compared to that of the star in Case 2.  In the final converged
time-step, this stellar model possesses a radius smaller than in
either of the standard Case 4 models.  Hence, although the envelope is
strongly formally dynamically unstable at this point, it is not clear
to us whether the star would experience dynamical instability during a
self-regulated spiral-in or eject the envelope during runaway
expansion.

The difference between Case 3 models with and without recombination energy release
was similar to the situation in Case 4. That is, 
the stellar expansion was slower than in both the bottom- and uniformly-
heated Case 3, however, envelope expansion still runs away after the 
hydrogen recombination front begins to propagate inwards.

The biggest difference between the cases with and without recombination
energy was in Case 2, in which recombination energy would have played a
stronger role compared to the other cases. When the recombination
energy was removed, the star did not run away and entered a self-regulated
spiral-in, reaching the same radius as in the Case 2 uniform
heating. However, the envelope is slightly more formally unstable than in the standard
Case 2, having $\langle\Gamma_1(m_{\rm bot}) \rangle=1.39$ at $\Delta
E_{\rm heat}^{\rm gross}=8.4\times 10^{46}$ ergs (which is the last
converged model).

We conclude that this definitively demonstrates that helium recombination energy  affects the
change of the stellar structure during a common-envelope spiral-in and
thereby alters the outcome of the common-envelope phase.

\begin{figure*}
\hskip-0.5cm\includegraphics[width=63.5mm]{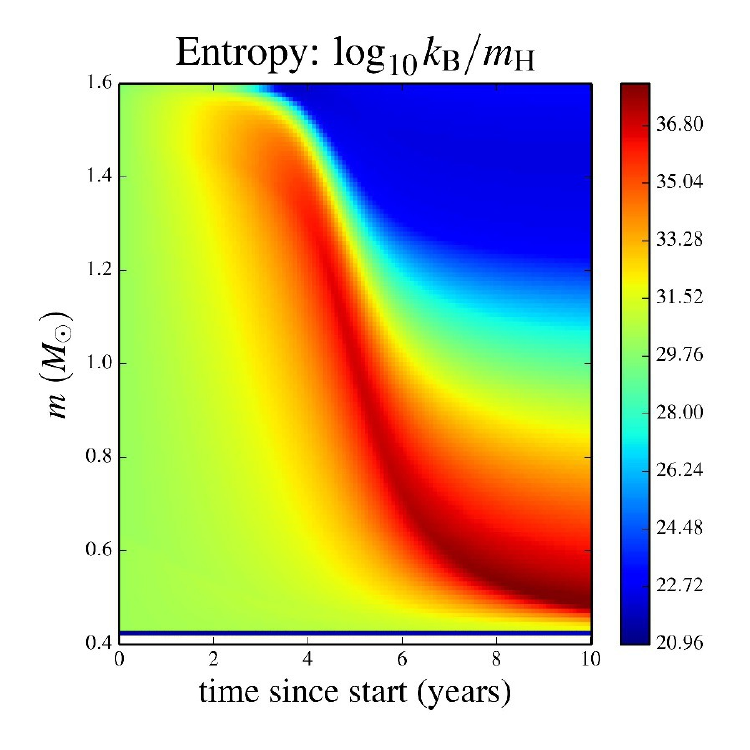}
\hskip-0.5cm\includegraphics[width=63.5mm]{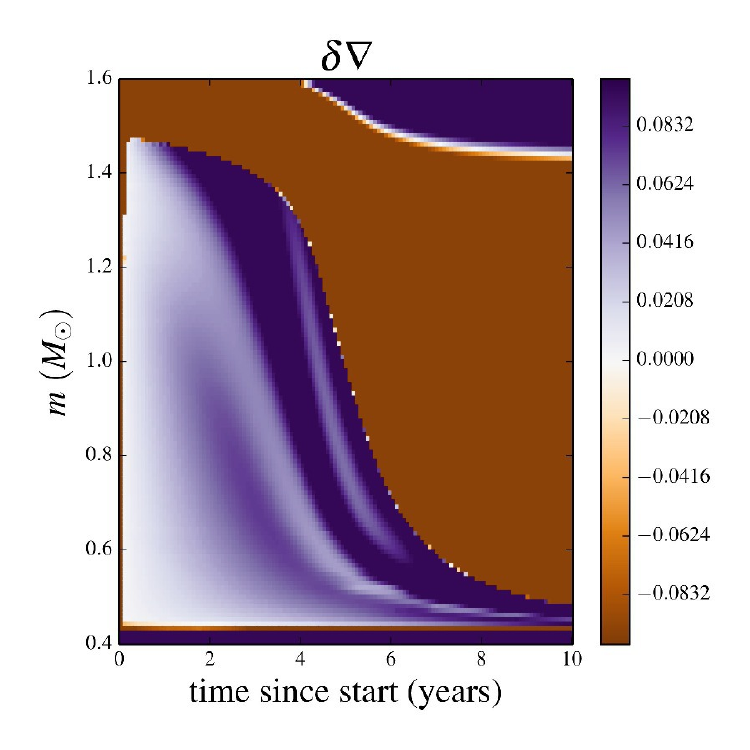}
\hskip-0.5cm\includegraphics[width=63.5mm]{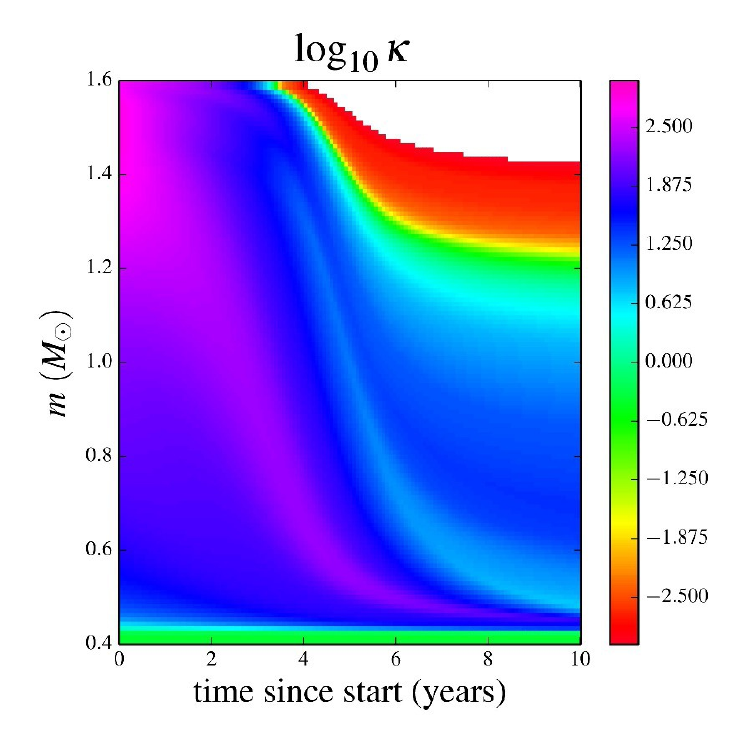}

\hskip-0.5cm\includegraphics[width=63.5mm]{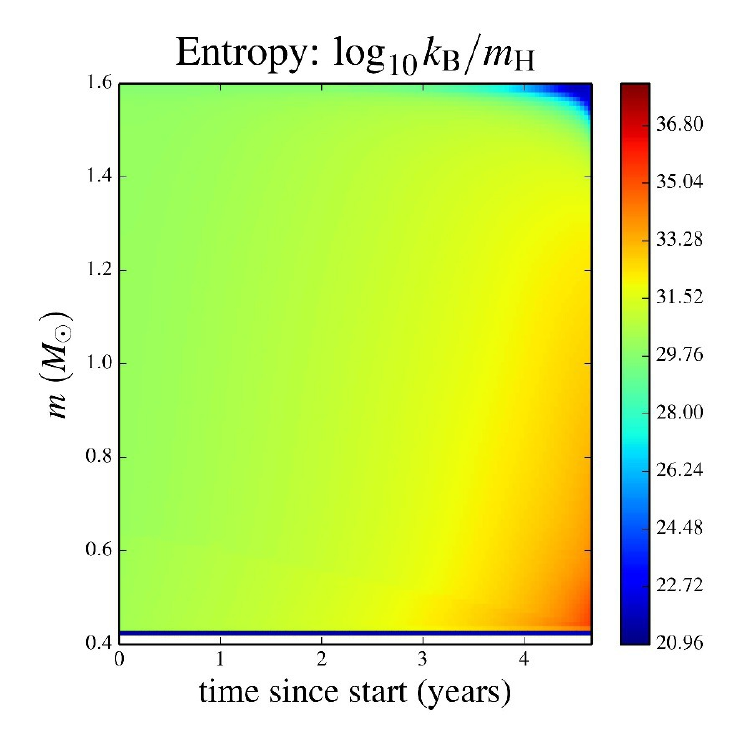}
\hskip-0.5cm\includegraphics[width=63.5mm]{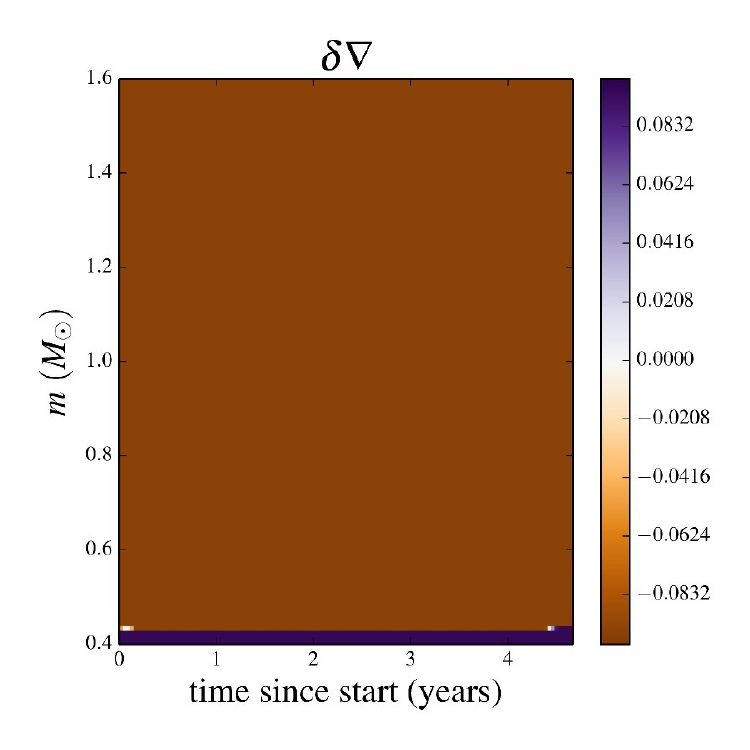}
\hskip-0.5cm\includegraphics[width=63.5mm]{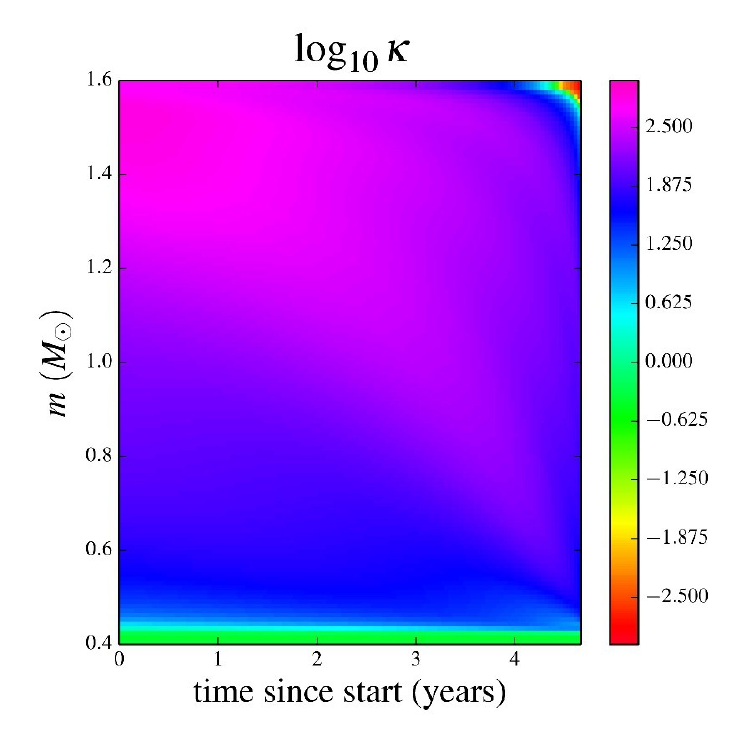}
\caption{The entropy (left panels), radiative and convective zones
(middle panels) and opacities (right panels) for the Case 4 heating
rate when adopting both uniform heating (the top panels) and bottom heating (the bottom
panels).  To clarify the presentation of the gradients, 
we truncated the value of $\delta \nabla$ at $-0.5$ in
convective regions and at 0.5 in radiative regions.}
\label{fig:2d_radcon} 
\end{figure*}

\subsection{Testing for the role of the location of the heating source}

We tested what would happen if heating was applied at the same specific
rate (i.e.\ the same $\varepsilon_{\rm heat}$ in erg g$^{-1}$), 
but when the location of the heating was changed. 
For this test we chose to shift the bottom boundary outwards by $-0.02, 0.05,
0.1$ and $0.15 {\rm{M}}_\odot$ compared to our standard bottom heating. 
We also chose the Case 2 rate of heating because this is
the case for which changing between uniformly-distributed and bottom heating
qualitatively alters the outcome from self-regulated to runaway expansion.
The first three changes of location made little difference to the results, but when
the inner boundary was moved outwards by $0.15 {\rm{M}}_\odot$, then 
the stellar expansion started earlier, increasing the star's surface
luminosity.  In this calculation, the model approached
the self-regulated solution (we will refer to this model as the MS015
version of Case 2 bottom heating).

\subsection{Testing different heating concentrations}

We also tested the change in outcome when the same total rate of
heating was concentrated within a different amount of mass.
Again we chose to compare to the Case 2 bottom-heating model, in which 
the heat was injected into a $0.1 {\rm{M}}_\odot$ layer.
In models with uniform heating,  the energy input is distributed over
almost 1.4 $\rm {\rm{M}}_\odot$ of envelope.
For this test we calculated examples where the Case 2 rate of energy
input was distributed in layers with masses of $0.025,0.05,0.2,0.4,0.6$ and $0.8
{\rm{M}}_\odot$, with inner edges located at the bottom of the convective envelope.
In each of the first four of those examples (i.e.\ with the specific heating rate
up to four times higher or four times lower than the baseline model),
the radius evolution did not differ at all from the standard Case
2 bottom-heating calculation. 

When the mass of the heated layer was increased to $0.6 {\rm{M}}_\odot$ (we will
refer to this model as the X6 version of Case 2 bottom heating), then the radius
evolution was slightly different -- with earlier expansion as the
effects of heating reached the surface earlier. This calculation also
appears to almost reach a
self-regulated state, but fails to do so (see more in \S\ref{sec_pul}). The last
converged model has a radius which is almost the same as in the
comparison calculations with
a smaller heated mass (including the standard Case 2 bottom-heating
calculation), but which is larger than the steady self-regulating
radius reached by the uniformly-heated calculation with the Case 2
energy input rate. 

When the energy input was distributed in $0.8 {\rm{M}}_\odot$ (which we shall
call the X8 version of Case 2 bottom heating), the initial expansion of the
star was just slightly slower
than for the uniformly-heated Case 2 and slightly faster than for the
standard bottom-heated Case 2 calculations.
At $\Delta E_{\rm heat}^{\rm gross}\approx5\times 10^{46}$ erg, the stellar
expansion overtook the uniform case and the model reached a nearly-self-regulated
radius at $\Delta \approx  E_{\rm heat}^{\rm gross} = 5.5 \times 
10^{46}$ ergs, slightly earlier than in the uniformly-heated
calculation. However, the self-regulation was not perfect, and the star
continued to expand very slowly. This expansion eventually ran
away at the same radius at which the runaway happened for the
calculations with smaller  $\delta m_{\rm heat}$.

\section{Discussion of physical processes}

\subsection{Convective and radiative regions and an entropy bubble}
\label{sec:entropy_bubble}

There is a striking difference between uniform and bottom heating in
the internal energy transport: the distribution of radiative and
convective zones. For uniform heating, convection halts. This is due
to a snowball effect which is triggered by heating throughout the
envelope. That heating leads to a temperature increase and,
accordingly, to a small decrease in opacities, decreasing the
radiative gradient. As radiation plays an increasing role in
transporting the energy and convection become relatively less
efficient (see examples on Figure~\ref{fig:2d_radcon}), an internal
radiative zone is created.  As a result, the local specific entropy
increases, creating an ``entropy bubble''.  The growth of this
entropy, and specifically the presence of a negative entropy gradient
(i.e.\ $ds/dm <0$) leads to the re-establishment of convection. Even
though the convection zone again extends further inside, it never
penetrates as deep as it had done before. At the surface, once
hydrogen recombination starts, each of our uniform-heating models
develops a radiative zone.

By contrast, the convective zone for bottom-heated models does not
change with time, except for the creation of a small surface radiative
zone in some cases.  We note that this implies that the structure of
the envelope during the evolution of the bottom-heated models remains
close to isentropic, but that this is not generally true for uniform
heating. We have checked that the absolute value of the adiabat in
bottom-heated models is almost constant in time. Hence the envelope
expansion caused by bottom-concentrated heating is close to adiabatic,
whilst uniform heating leads to a strongly non-adiabatic envelope
expansion.  The exception is Case 5 of uniform heating, for which the
specific entropy is also mostly uniform throughout the envelope, most
likely because the duration of the evolution is too short for it to
change.  In \S\ref{sec_rec} we argue that this difference in
``adiabatic'' versus ``non-adiabatic'' envelope expansion is
significant in understanding the differing usefulness of recombination
energies.

\subsection{The location of the photosphere}

Once the hydrogen recombination front begins to move inwards from the
surface, the opacities of the outer layers drop dramatically.  
It is these layers with recombined hydrogen that become radiative.
Sometimes as much as the outer $0.2 {\rm{M}}_\odot$ 
can contain neutral hydrogen and be radiative. Moreover, a significant
fraction of that mass can be optically very thin and is likely
located above the photosphere of the star. 

This is qualitatively similar to the expected structure of a red
supergiant, with a cool extended atmosphere and convection starting to
play a role in the energy transport only at large optical depths
\citep[see ][]{1969AcA....19....1P}. 
It is not intuitively clear whether any simple photospheric
condition can be used to model such stars. In the past, it was even argued
that a zero boundary condition at $\tau=0$ with 
$L = 8\pi \sigma R^2 T_0^4$ should be used \citep[][see this paper for
how to obtain the radiative temperature gradient 
in the atmosphere with this boundary condition]{1969AcA....19....1P}. 
Here $T_0$ is the temperature at $\tau=0$. 
However, it has also been shown that, for these
optically thin atmospheres, the chosen boundary condition
does not affect the solution as long as the atmospheric opacities are
sufficiently low \citep{1969AcA....19....1P}. 
Although the stellar structures we obtain are no more inconsistent than
standard stellar models of red supergiants, we realise that it may be 
interesting for future studies to investigate how modified atmospheric
models affect our results.

The work of \citet{Ivanova+2013Science} suggested that the
recombination of hydrogen in a large fraction of the envelope mass did
not occur until \emph{after} the envelope was ejected. This may well
still be consistent with our results here if the remainder of the
envelope is rapidly ejected.  We note that, as discussed in the
Introduction, our neglect of the dynamical terms seems likely to lead
to systematically \emph{later} ejection in these calculations than in
reality.  So perhaps in reality the hydrogen recombination front has
less time to propagate inwards before envelope ejection.

\subsection{Recombination energies} 
\label{sec_rec}

\begin{figure*} 
\includegraphics[width=85mm]{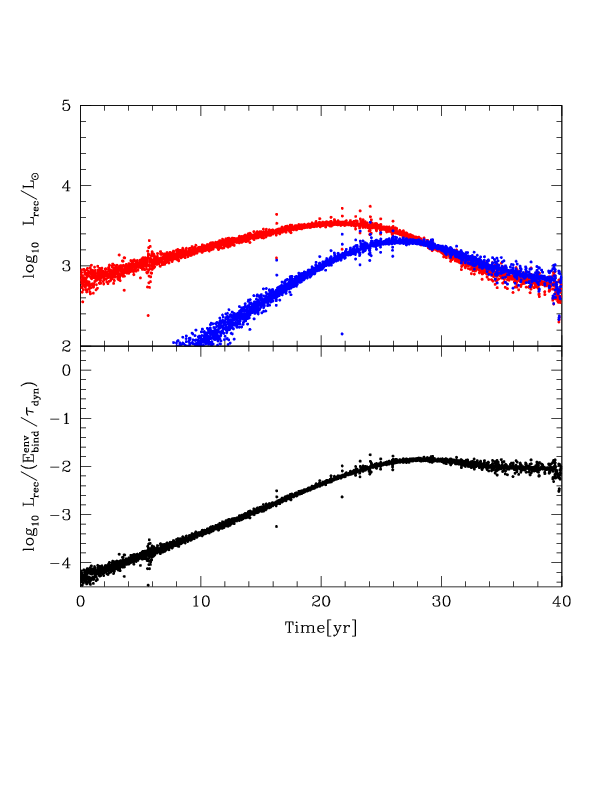}
\includegraphics[width=85mm]{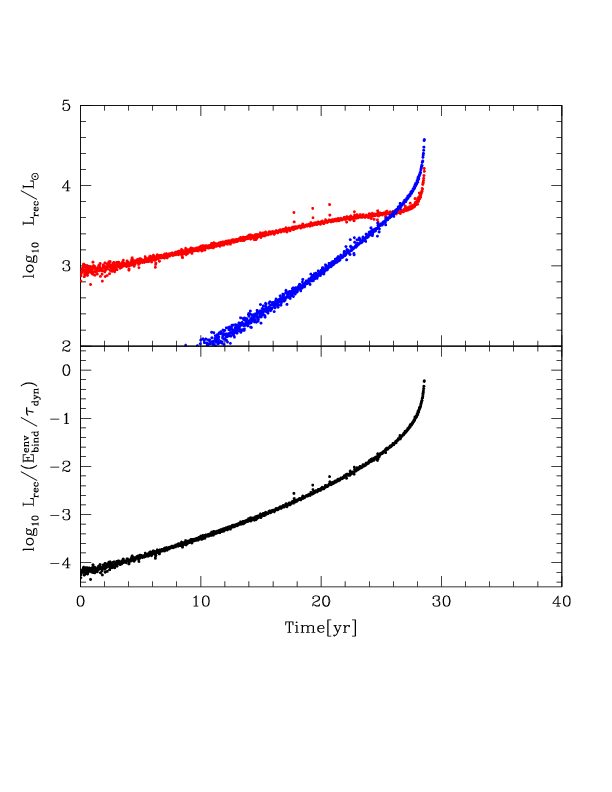} \vskip-2cm
\includegraphics[width=85mm]{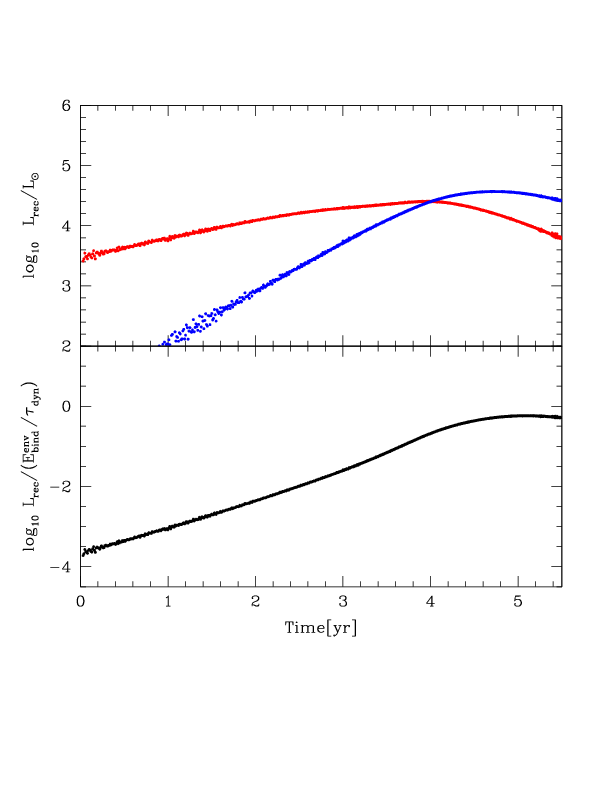}
\includegraphics[width=85mm]{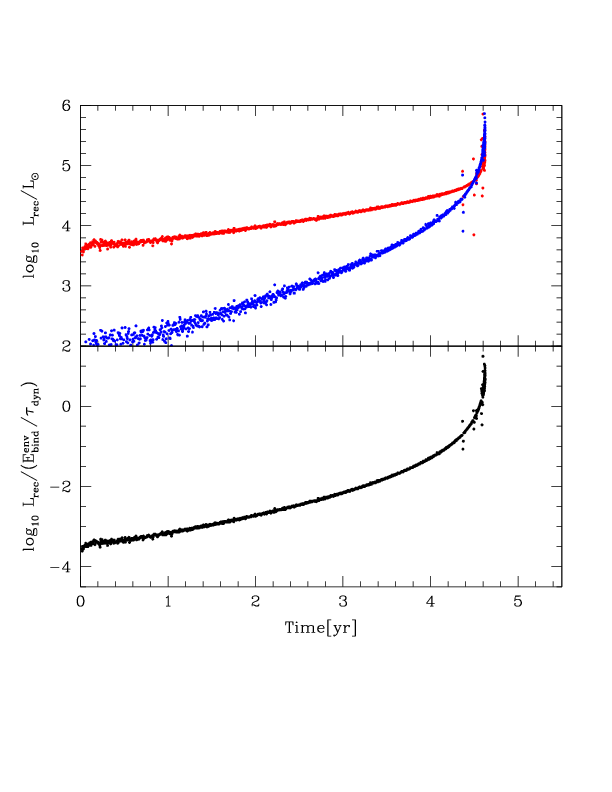} 
\caption{The role of recombination luminosities in Cases 2 (upper pairs of plots) and 4 (bottom
pairs of plots), uniform heating (left) and bottom (right) heating. The top panels in
each case show the energies provided by hydrogen $L_{\rm rec,H}$ (blue)  and
helium $L_{\rm rec,He}$  (red) recombination. The bottom panels 
show the ratio between the total recombination luminosity $L_{\rm
rec}=L_{\rm rec,H}+L_{\rm rec,He}$ and the ``dynamical'' luminosity -- 
which we define as the current binding energy of the envelope divided
by the envelope's current dynamical time. 
Note that all of the recombination luminosities are
derived quantities which were not used during the evolutionary calculations.
The noise is mainly due to the way in which the changes in the
ionisation states between models were calculated during the post-processing, with
larger noise when the models were re-meshed by mass.}
\label{fig:rec_lum} 
\end{figure*}

We introduce the recombination luminosities:

\begin{equation} 
L_{\rm rec, H}=\int_{M} \varepsilon_{\rm rec, H} dm ,
\end{equation}

\noindent and

\begin{equation} 
L_{\rm rec, He}=\int_{M} \varepsilon_{\rm rec, He} dm  . 
\end{equation}

\noindent 
Here $\varepsilon_{\rm rec, H}$ and $\varepsilon_{\rm rec,
He}$ are local specific energy generation rates due to the recombination of hydrogen and
helium (respectively).

We also define the ``dynamical'' luminosity as the ratio of the
binding energy of the envelope to the dynamical time-scale of the star,
$\tau_{\rm dyn}$:

\begin{equation} 
L_{\rm dyn} = \frac{E_{\rm bind} (m_{\rm bot})}{\tau_{\rm dyn}}   . 
\end{equation}

\noindent For this estimate, we use $\tau_{\rm dyn} = \sqrt{R^3/GM}$.

In Figure~\ref{fig:rec_lum} we show the recombination luminosities 
$L_{\rm rec, H}$ and $L_{\rm rec, He}$, as well as the ratio of the total
recombination luminosity $L_{\rm rec} = L_{\rm rec, H} + L_{\rm rec,
He}$ to the star's dynamical luminosity. We find that in all the models
where the envelope expansion runs away, the total recombination luminosity exceeds
the dynamical luminosity. This helps to explain why those stars can not remain in
equilibrium.  In this ``runaway'' regime, the recombination luminosity is
predominantly provided by hydrogen recombination, which increases
sharply just before the expansion runs away.  
The rapid growth of $L_{\rm rec, H}$ appears to drive a similar
increase in $L_{\rm rec, He}$.  Before the onset of this runaway
behaviour,  the evolution of $L_{\rm rec, He}$ is 
almost independent of how the heating regions are
distributed, only on the total energy input.

The maximum $L_{\rm rec, H}$ with which we see the star enter a
self-regulated state is $\log L_{\rm rec, H}/{\rm{L}}_\odot = 4.6$ (the
Case 4 uniform heating). In any model we have calculated in which 
$\log L_{\rm rec, H}/{\rm{L}}_\odot > 4.6$, we find that the envelope 
expansion starts to run away. The connection between
recombination luminosity and the mass of stellar material that is recombining
is:

\begin{equation} 
L_{\rm rec, H} \approx 2.1 \times 10^{5} \ X   
\dot M_{\rm rec, H} [{\rm{M}}_\odot {\rm yr}^{-1}] 
\end{equation}

and

\begin{equation} 
L_{\rm rec, He} \approx 3.2 \times 10^{5} \ Y  
\dot M_{\rm rec, He} [{\rm{M}}_\odot {\rm yr}^{-1}]. 
\end{equation}

\noindent  Where $X$ and $Y$ are the mass fractions of hydrogen and
helium, and $\dot M_{\rm rec, H}$ and $\dot M_{\rm rec, He}$ are the rates
at which hydrogen and helium recombine, in M$_\odot$ yr$^{-1}$.
When $\log L_{\rm rec, H}=4.6$, as in the case described above, 
the recombination of hydrogen proceeds at a rate of $\sim
0.3 \rm {\rm{M}}_\odot yr^{-1}$.  Therefore the associated recombination
front is moving inwards through the envelope on a time-scale almost as
short as the dynamical time-scale of the expanded star.  We note that
for these ``self-regulated'' stellar  structures, or when the
expansion starts to run away, the hydrogen recombination zones are at
an optical depth significantly more than one. 
Because the recombination front is not thin, there is no
  single unambiguous way to define the depth of the recombination zones.
The simulations show
$\varepsilon_{\rm rec}<10^3 {\rm erg~g{^{-1}}~s^{-1}}$ when the ionisation 
fraction is below 0.1, which is at $\tau\la 500$. We also see 
$\varepsilon_{\rm rec}\la 10^2 {\rm erg~g{^{-1}}~s^{-1}}$ at $\tau \la 50$. 
Hence is it plausible that a significant fraction of the hydrogen
recombination energy is used to expand the star in these situations.

Helium recombination energy is usually almost fully utilized for
driving envelope expansion. Only at each extreme of the heating rates
we considered, a large fraction of this energy was not released by the
end of the simulations.  For Case 1, the rate of energy input is so
small that $\dot M_{\rm rec, He}$ is also very small. For Case 5 with
uniform heating, the expansion of the stellar surface runs away before
the inner layers expand sufficiently to start recombining
(i.e.\ ``dynamical heating'' of the surface).

But why does the expanding star sometimes survive hydrogen
recombination whilst for other models the expansion runs away?
Consider the Saha equation for pure hydrogen, where we denote
the ionisation fraction of hydrogen as $y=${H{\small II}/H}:

\begin{eqnarray} 
\frac{y^2}{1-y} &=& {4 \times 10^{-9}} \left( \frac{T}{\rm K}
\right)^{3/2} \left( \frac{\rm g~cm^{-3}}{\rho} \right)
\exp{\left( - \frac{1.58\cdot 10^5 {\rm K}}{T} \right)} \nonumber \\
& \equiv & F(\rho,T).
\label{eq:saha}
\end{eqnarray}

\noindent Here $T$ is the temperature in $\rm K$, and $\rho$ the density
in $\rm g~cm^{-3}$. The left-hand side of the equation,
$y^2/(1-y)$, monotonically changes with $y$. When the right-hand side of
the equation ($F(\rho,T)$) decreases, the ionisation fraction $y$
also decreases. Hydrogen recombination starts with decreasing $F(\rho, T)$,
 and hydrogen becomes half recombined when $F(\rho,T)=0.5$.
While we anticipate that in a complete equation of state (EOS),
$F(\rho, T)$ has a more complicated form due to the presence of helium
and free electrons from other elements, the dependence on the
temperature and density for hydrogen ionisation fractions is broadly
determined by the simple Saha equation for pure hydrogen as above, 
since helium ionization is very small until hydrogen is almost fully ionized, and
hence helium does not provide many free electrons. The density and
temperature, through $F(\rho, T)$, therefore determine the
degree of hydrogen ionisation.

For this analysis, we will assume a simple power-law EOS of the form
$T\propto \rho^x$. Here, $x=2/3$ would correspond to adiabatic changes
of a monatomic ideal gas, and so the term $T^{3/2}/\rho$ in $F(\rho, T)$ is
constant in case of an adiabatic change. If the entropy of the material
increases, the change is non-adiabatic, and this situation could be
described by using $x<2/3$. Plasma with a higher entropy
has a larger $F(\rho,T)$ and is less recombined.

The rate of recombination energy release from hydrogen is proportional to the change in
ionisation fraction: $\varepsilon_{\rm rec, H}\propto - dy/dt$. From 
 Eq.\ (\ref{eq:saha}) we obtain:

\begin{eqnarray} 
\frac{dy}{dt} &=& \frac{(y-1)^2}{(2-y)y}  \left (\frac{3}{2}x- 1 
+ x \frac{1.58\cdot 10^{5} {\rm K}}{T}\right ) \frac{d \ln \rho}{
d t} F(\rho,T) \nonumber \\ 
& \approx & {6.3\times10^{-4}} 
\frac{(y-1)^2}{(2-y)y}  
\left( \frac{T}{\rm K}
\right)^{1/2} \left( \frac{\rm g~cm^{-3}}{\rho} \right) \nonumber  \\
&~& \times \exp{ \left( - \frac{1.58 \times
10^{5} {\rm K}}{T} \right)} x  \frac{d \ln \rho}{ d t}  . 
\label{eq:erec_saha}
\end{eqnarray}
We can now use the framework described above to examine the differing 
outcomes of envelope expansion (as parametrized by decreasing
density). We have previously argued that the two relevant limiting cases
are adiabatic expansion and expansion with entropy increase  (see \S
\ref{sec:entropy_bubble}).  For the same expansion rate ($d \ln \rho/
d t$),  at every instant, the temperature in the adiabatic case
will be higher. The value of $x$ is also higher for adiabatic
expansion than for  expansion with entropy increase. Therefore
Eq.~\ref{eq:erec_saha} helps to explain why the recombination
energy  release occurs at a higher rate for adiabatic expansion. 
Of course, once recombination starts, the plasma does not 
continuously move along its adiabat, and so this discussion can only
indicate the characteristic behavior in a very simplified way. 

We now compare stellar envelopes with the same {\it radius} and for
the  same  heating  rate.   Note that these
stars  may  have received a different total heating energy input from
each other when they  have the same radius, but this difference is
small before the envelope expansion
settles to a self-regulated solution or starts to run away.

For bottom-heating cases, due to the ongoing convection (see  \S
\ref{sec:entropy_bubble}),
 the envelopes
are almost ideally isentropic both in space (throughout the
envelope) and in time (see Fig.~\ref{fig:2d_radcon}). 
On the other hand, the cases with uniformly-distributed
heating form strongly non-isentropic envelopes, also both in space and
in time.
Hence, when stars with the same radius and which are heated at the
same rate are compared, 
$F(\rho, T)$ values in the bottom-heated cases are overall smaller
than in stars with significantly non-isentropic envelopes.
In addition, the $F(\rho,T)$ derivative with respect to mass is smaller in isentropic
envelopes -- it is more nearly constant over a larger mass range within the
envelope.
Because hydrogen recombination follows $F(\rho,T)$,  
at any instant in an isentropic envelope: (i) hydrogen is recombining in a larger
range of mass and (ii) the envelope is overall more recombined 
than in a non-isentropic envelope with the same radius.

Whilst this analysis is very simplified, this difference in behavior 
is present in our simulations that use complete EOS. We see a noticeable
difference in $F(\rho,T)$ values and profiles when comparing these two
types of envelopes. The difference can be obvious when the stars have
only expanded to $200 {\rm{R}}_\odot$, well before the expansion starts to run away.

For envelopes with the same radius, we therefore expect that the rate
of recombination energy release is greater in isentropic envelopes. 
Accordingly, we see higher local values of  $\varepsilon_{\rm rec, H}$
at every single moment in the calculations of isotropic envelopes
(i.e.\ those with bottom heating)  than
in the non-isentropic envelopes (i.e.\ those with uniform heating).

We estimate the local rate of energy input which is capable of
disturbing  local hydrostatic equilibrium as $\varepsilon_{\rm pot} =
GM/(r\tau_{\rm dyn}(r))$ (comparable to the global dynamical luminosity $L_{\rm
  dyn}$, as defined earlier).  This provides a natural scale to which
we can compare the local rate of energy release from recombination, 
$\varepsilon_{\rm rec}$.  Comparing two stars
heated at the same total rate and with the same stellar radius, we
find that models with adiabatic envelopes have a significantly higher  $\varepsilon_{\rm
rec}/\varepsilon_{\rm pot}$ than those with non-adiabatic (``entropy-bubble'') envelopes. 
For those adiabatic envelopes, the ratio $\varepsilon_{\rm rec}/\varepsilon_{\rm pot}$
can be as high as 10.  When  $\varepsilon_{\rm rec}/\varepsilon_{\rm
pot}\gg 1$ in a substantial part of the envelope, then 
hydrogen recombination provides ``dynamical''
heating,  and the envelope expansion runs away. (Recall that
isentropic envelopes tend to show significant recombination over a
relatively large range of masses, which also helps to produce this outcome.)

From all of the above, we conclude the the radiative zone which
develops in uniformly-heated envelopes is key in
slowing down the overall rate of
recombination and consequently prevents runaway expansion.

\subsection{Super-sonic and Super-Eddington convection and mass loss}
\label{sec:SuperSonicConv}

\begin{figure}
\includegraphics[width=90mm]{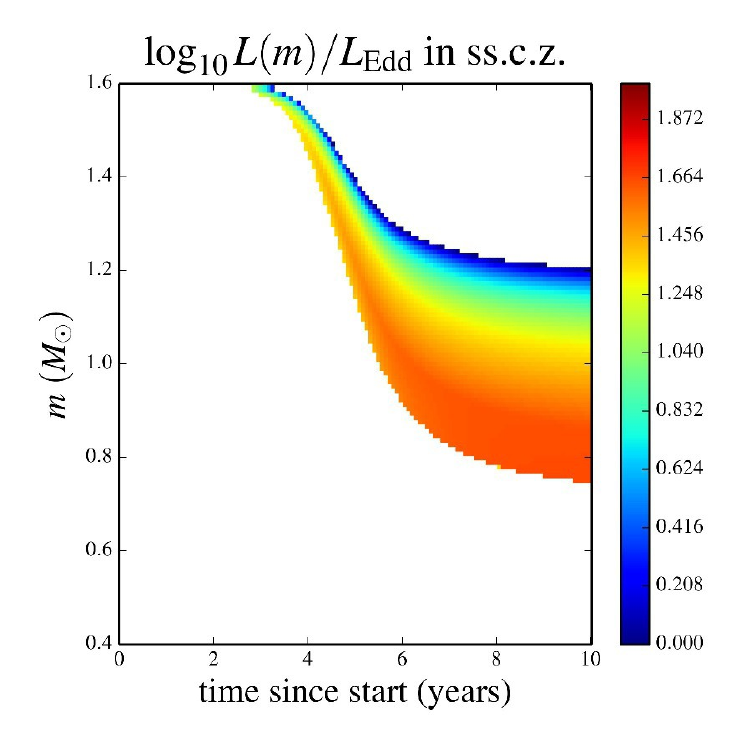}
\caption{The ratio of the local luminosity to local Eddington
luminosity, shown only for those convective regions where subsonic
convection cannot carry the energy flux anymore for the
uniform heating Case 4.} 
\label{fig:2dssconv}
\end{figure}

While in the uniform case the luminosity increases towards the
surface, in the bottom heating the luminosity decreases towards the
surface. Within radiative zones, as expected, the luminosity is always
locally sub-Eddington (here the local Eddington luminosity was
calculated using local opacities). However, for bottom
heating, the local luminosity in convective zones can be substantially super-Eddington.

When the internal heating becomes sufficiently high, the outwards energy
flux may also exceed the amount which normal sub-sonic convection described
by the mixing length theory would be capable of carrying (see
Figure~\ref{fig:2dssconv}). A similar problem is known to occur sometimes in
very massive stars when neither sub-sonic convection nor radiation
 can carry all of the required energy flux
\citep[][]{2012MNRAS.423L..92Q,2014ApJ...780...96S}. If that joint
condition on radiative and convective energy transport is met when the
star also has a surface radiative zone, it has been argued that
wave-driven mass loss will occur \citep[][]{2012MNRAS.423L..92Q,2014ApJ...780...96S}.
The rate of this mass loss has been predicted to be as
large as $1 {\rm{M}}_\odot~{\rm yr^{-1}}$
\citep{2012MNRAS.423L..92Q,2014ApJ...780...96S}. While this situation
does not occur in all the models we considered, it did happen for Cases
2,3,4 (uniform heating), and for three variations of the Case 2 bottom
heating model (X6, X8 and MS015). The luminosity at the top of the
convective zone is usually up to a few times larger than the local
Eddington luminosity, suggesting that the heated envelope may experience
substantial wind mass loss.

\subsection{Growth of instability}

\begin{figure*}
\includegraphics[width=58mm]{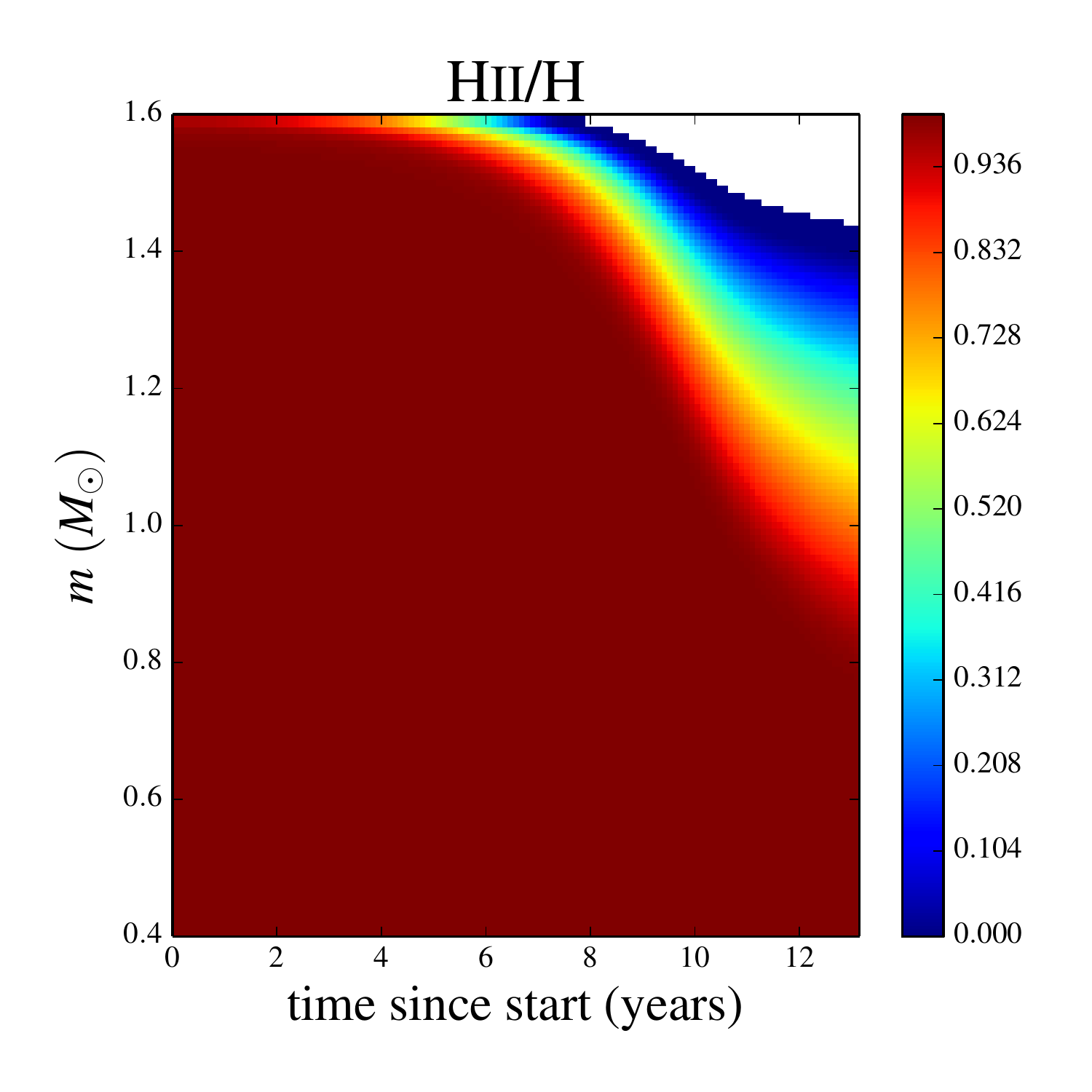}
\includegraphics[width=58mm]{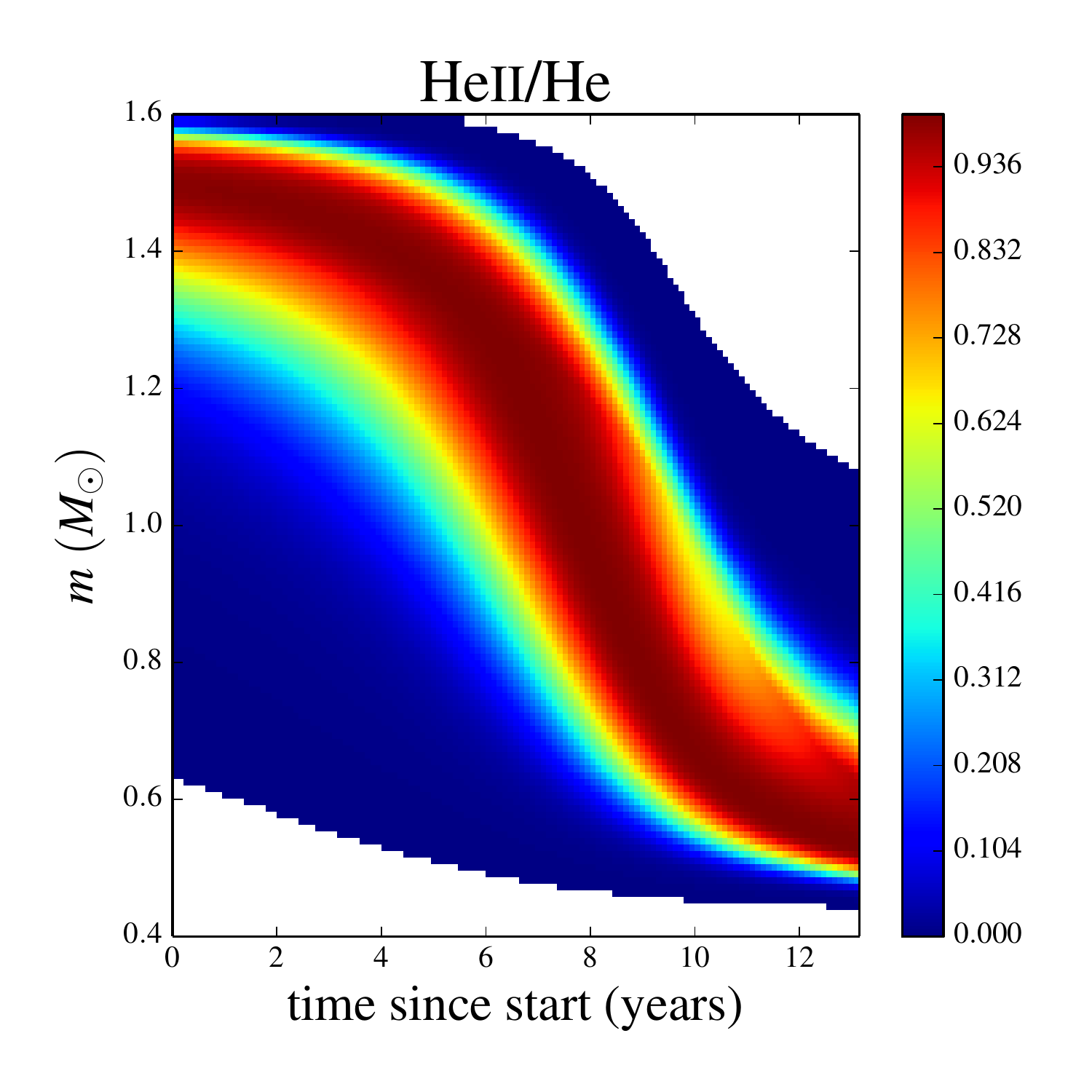}
\includegraphics[width=58mm]{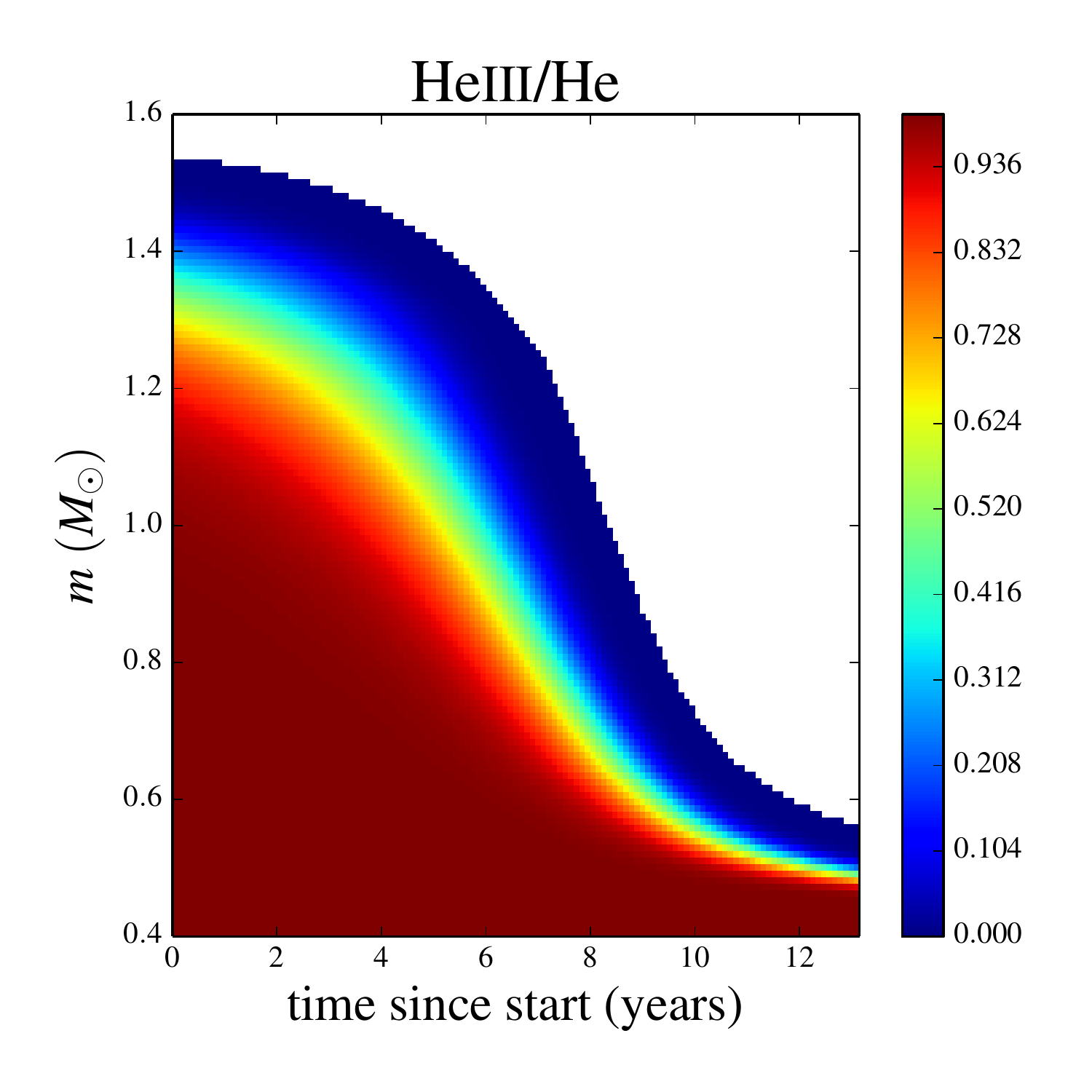}
\includegraphics[width=85mm]{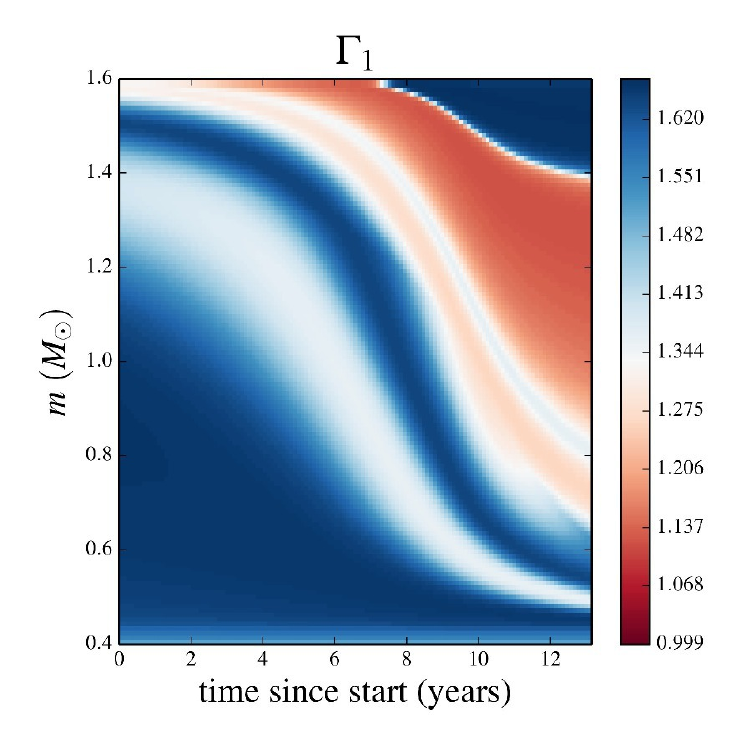}
\includegraphics[width=85mm]{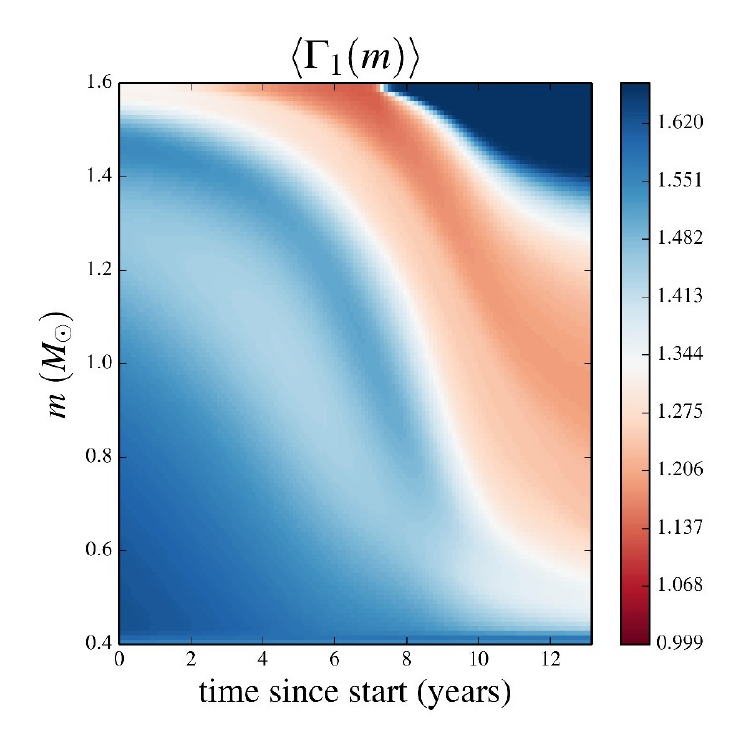}
\caption{Ionisation fraction of hydrogen and helium,  local $\Gamma_1$
and integrated from the surface pressure-weighted volume-averaged 
$\langle \Gamma_{1}(m) \rangle$ in the Cases 3  uniform heating .}
\label{fig:2d_instab} 
\end{figure*}

Hydrogen recombination plays the most important role in the overall
formal instability of the envelope -- in the region of partial hydrogen
recombination, $\Gamma_1$ is minimal (see Figure~\ref{fig:2d_instab}).
The integral of this quantity inwards from the surface, $\langle \Gamma (m) \rangle$,
implies that most of the envelope in almost all of the models we calculated  
is formally dynamically unstable
(the only exception is that of Case 5 with uniform heating -- ``dynamical''
heating).  We stress that we are not
referring here to the importance of hydrogen recombination to the
\emph{energetics} of ejection, but to the value of $\Gamma_1$.

\subsection{Pulsations}
\label{sec_pul}

\begin{figure}
\includegraphics[width=90mm]{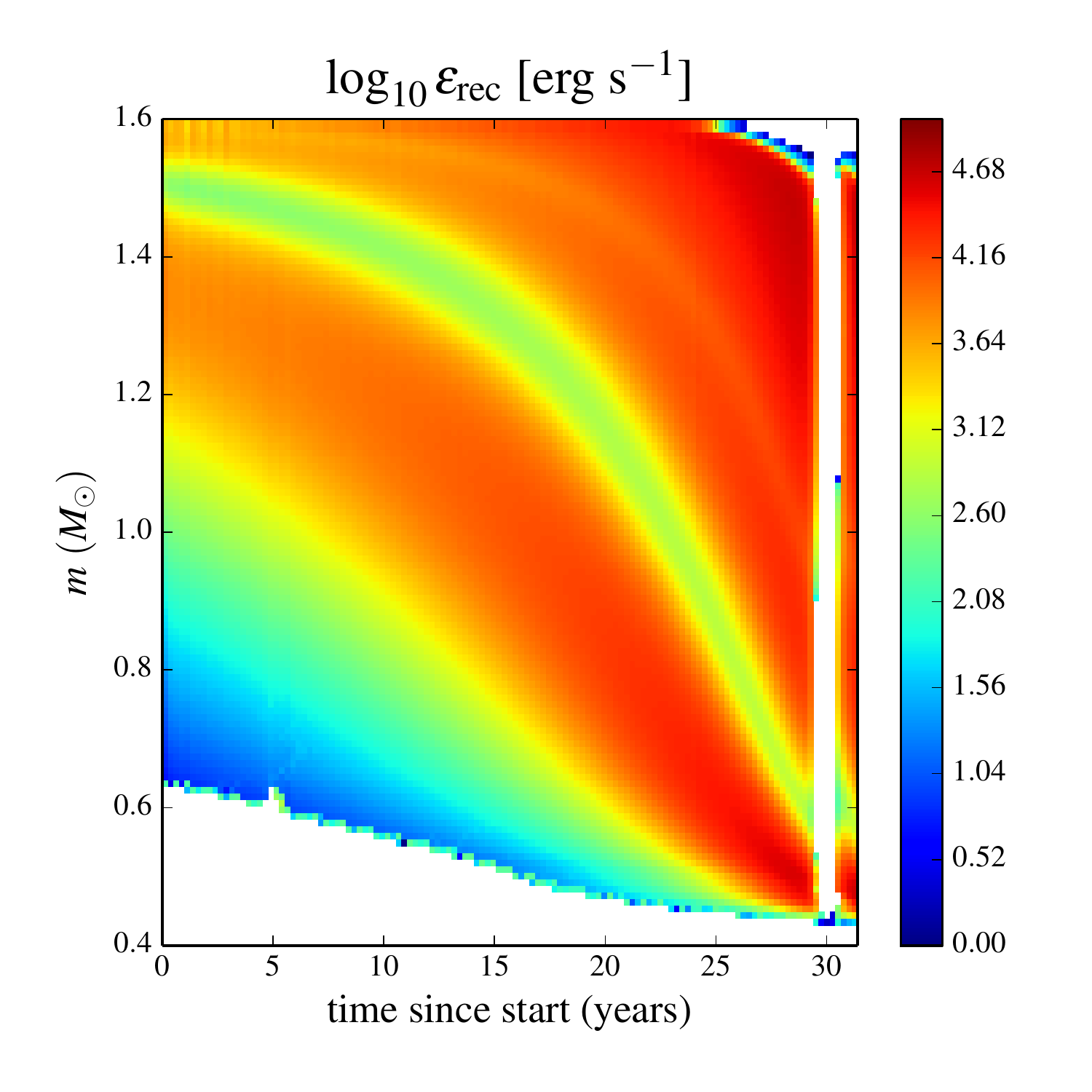}
\caption{The rate of recombination energy release during the pulsations displayed by model
  X6. The blank vertical region at roughly 30 years occurs during the
contraction phase of the pulsation, during which matter is re-ionised and
absorbs energy.}
\label{fig:enrec_puls} \end{figure}

While we anticipate dynamical instability of the envelopes in
general, we have also encountered {\it relaxation pulsations} in our calculations. The
pulsations take place at the moment when the stellar radius is
about to reach self-regulation (in terms of radiating away the heating
luminosity). 
If the star expands too fast at this point, the expansion
runs away. 
If the star is expanding sufficiently slowly,  the transition from 
expansion to a self-regulated state takes
place without noticeable pulsations. For intermediate rates of
expansion, the stellar envelope first overshoots the
self-regulated radius and then contracts back below the self-regulated
radius. 
Obvious pulsations were first noticed in cases X6, X8 and MS015.  
The contraction phase of each pulsation is accompanied by {\it
  re-ionisation} of the material all the
way towards the bottom of the envelope, and therefore also involves 
helium re-ionisation (see Figure~\ref{fig:enrec_puls}), which absorbs
energy.

\subsection{The core and its energy}

It has previously been unclear whether the response of the core during
CEEs might significantly affect the energy budget of envelope ejection
\citep[see, e.g.,][]{Ivanova+2013Review}. 
In all of our simulations, the total binding energy of the core became
only slightly {\it less negative}.
We presume that this energy was taken from nuclear energy
release, although we cannot be sure. 
However, the accumulated difference in binding energy is roughly
$0.05\times 10^{46}$ and hence is small compared to the other terms in
the energy budget which are discussed elsewhere in this paper.

\subsection{Are these results unique to this particular choice of the
core mass?}

\label{sec:common}
We have re-calculated several examples using different initial
stellar models, specifically ones with a smaller core mass (since it
has broadly been expected that recombination energy will be more
significant for CEE involving giants with larger radii). 
We applied similar prescriptions in terms of the heating rate per unit
of time and the mass distribution 
(i.e.\ with the heating confined to the bottom $0.1~{\rm{M}}_\odot$ or
spread through the entire envelope). We obtained strikingly similar
results. The outcomes were qualitatively the same, in that the same heating
rates resulted in either runaway or self-regulated outcomes as
in our standard models, albeit at different values of $E_{\rm heat}^{\rm
  gross}$. In some respects they were also quantitatively very similar,
since the steady-state luminosities and radii
were roughly the same as in our base case (they were slightly lower by
an amount which corresponds to the lower nuclear luminosity of the
initial model). For such smaller core masses the initial binding energy of the
envelope is significantly more negative. The final outcomes for these
models take place  at slightly higher values of $E_{\rm heat}^{\rm
gross}/E_{\rm bind}$ than in the comparable standard models.

\section{Conclusions}

We have studied simplified models of the phase of CEEs which is
expected to follow the initial dynamical plunge.
These calculations suggest that 
such heating could produce two types of outcome:

\begin{itemize} 

\item ``runaway'' -- the envelope expansion
  accelerates until it starts to escape on its current dynamical time;

\item ``self-regulated'' -- the envelope expands enough to radiate away
the heating luminosity. However, even in these cases most of the envelope becomes
formally dynamically unstable. 
\end{itemize} 

Which of these outcomes occurs is determined not only by
\emph{how much} heating energy is provided to the star, but is
strongly dependent on \emph{where} and \emph{at which rate} the
heating energy is provided.

In all the cases we considered, the envelope either reached runaway
expansion or a formally dynamically unstable state of self-regulation
after receiving \emph{less} net heating energy than the initial
binding energy of the envelope (where we define this
binding energy as gravitational potential energy plus internal
thermal energy but without the recombination energy terms).  This
confirms that the release of recombination energy can be energetically 
important to common-envelope ejection.

To quantify and illustrate the approximate importance of recombination, we define the
efficiency $\eta_{\rm min}$ -- the ratio of the \emph{gross} heating luminosity
to the binding energy (again including only gravitational and thermal
terms in $E_{\rm bind}$, without recombination energy).  
Then the standard energy formalism can be rewritten as

\begin{equation} 
\alpha_{\rm CE} \Delta E_{\rm orb} = \eta_{\rm min} E_{\rm bind} \ . 
\end{equation}

When some of the heating occurs close to the surface, the star
can enter an energetically``self-regulated'' state in which the
expanded star radiates away all the additional energy input.
However, even during this
``self-regulated'' stage, the envelope is formally dynamically
unstable.\footnote{Whilst this apparent self-regulation of
  radius expansion is not caused by the same mechanism
as the self-regulation in \citet{mmh79}, we note that our results are
directly relevant to that phase of slow spiral-in. In
particular, we consider it very likely that the solutions of
\citet{mmh79} are close to formal dynamical instability.}  
As much as 90 per cent of the envelope can be dynamically unstable for
$\eta_{\rm min}\approx 75$ per cent. These
envelopes may also experience strong mass-loss due to wave-triggered winds
(see \S \ref{sec:SuperSonicConv}), or from Mira-type winds.

Importantly, we find that a higher heating rate makes the same amount of input
energy more effective. Less massive secondaries are expected to plunge-in
faster, and a smaller fraction of the gravitational energy release
seems likely to be dissipated in the outer parts of the envelope. 
This suggests that the orbital energy release from relatively less massive
secondaries might be more effectively used in removal of the
envelope \citep[see also][]{podsi01}.

We note that this argument qualitatively fits the
inference from observations that the ejection efficiency grows when the companion is
less massive \citep{2011MNRAS.411.2277D}. However, we stress that
\cite{2011MNRAS.411.2277D} suggested that a low-mass companion would
orbit for longer and that the longer time-scale would allow a giant to use its
thermal energy, while our results suggest that a shorter time-scale for
the spiral-in leads to a higher efficiency in the use of recombination
energy.   

Further studies of these relative ejection efficiencies will require the use of
realistic frictional luminosity distributions. It might be that in
some cases viscous heating occurs through a large fraction of a
differentially-rotating envelope, whilst in others the heating is
dominated by dissipation close to the secondary. For example, the
in-spiral of relatively more massive secondaries
may tend to generate more broadly differentially-rotating
envelopes. If so, the envelope heating may be comparatively more
widely distributed during the in-spiral of relatively more massive
companions. Our calculations indicate that this would further 
reduce the envelope ejection efficiency
of more massive companions, in addition to the differences 
caused by different in-spiral time-scales.

The ionisation state of hydrogen plays two distinct roles in the
outcomes of our calculations. Most clearly and generally, it is most important
in controlling for how much of the envelope the value of
$\langle \Gamma (m) \rangle$ is low enough to indicate formal
dynamical instability.  In addition to that, our models sometimes
show hydrogen recombination fronts which produce dynamically-dominant
rates of heating (see \S \ref{sec_rec}).  Once hydrogen recombination is triggered in
a dynamical mode, we speculate that this may be capable of removing the entire
envelope, although our models cannot yet confirm this.
Such dynamical hydrogen recombination fronts can be triggered in our
calculations for $\eta_{\rm min}\approx 65$ per cent.

We stress that higher heating rates lead to lower $\eta_{\rm min}$. 
Heating which is confined to the lower half of the envelope -- which
seems likely for low-mass companions -- may trigger dynamical
hydrogen recombination at heating luminosities as low as $2\times
10^{45}$ erg yr$^{-1}$. For our initial stellar model, this heating
rate could be provided by the spiral-in of a
0.3 ${\rm{M}}_\odot$ companion over a time-scale as long as 500 years.

We expect that first-ascent giants with more massive cores
(i.e.\ stars in which the potentially available recombination energy is a larger
fraction of the initial binding energy of the envelope) would require even smaller
 $\eta_{\rm min}$ to produce each of the qualitative outcomes. 
However, we do not expect the change in $\eta_{\rm min}$ with
stellar radius (or core mass) to be linear.   

We have argued that differences in the progress of hydrogen
recombination are primarily responsible for determining 
which of the qualitatively different outcomes occur. These differences
are in turn a consequence of the entropy profiles and
convective-radiative structures of the envelopes (see \S
\ref{sec:entropy_bubble} and \S \ref{sec_rec}).
However, we have not been able to easily parametrize 
the importance of hydrogen recombination, either to the energetics or
the stability of envelope ejection. 
The total energy released from
recombination of hydrogen by the end of the simulations varied from 1 per cent to
60 per cent of the initial energy reservoir. Nonetheless, we consider that 
the most important effect of hydrogen recombination effect is the way
in which the ionisation state controls the formal dynamical stability
of the envelope. This dynamical destabilization takes place when
hydrogen is still almost fully ionised. 
The understanding of the role of hydrogen recombination
requires further study.

On the other hand, it is clear that in most cases, independent of
both the location and amount of heating, and also independent of the
qualitative outcome of the calculations (i.e., ``runaway'' or
``self-regulated''), about 90 per cent of the
recombination energy which was initially stored in ionised helium 
is used to expand the envelope. This apparently robust result
suggests that it is safe to include helium recombination as an additional
energy source in the energy budget for common-envelope ejection when CEEs 
proceed past the dynamical plunge-in stage.

Our results therefore support the use of this recombination energy when
CEE outcomes are estimated by use of an energy formalism. 
However, in order to start helium recombination, we expect that the
companion should already have plunged deep inside the envelope, even 
for envelopes which could be described as initially ``unbound'' if
their recombination energy were taken into account (e.g.\ AGB stars).
So our results should not be taken to support the notion that 
release of recombination energy can lead to envelope ejection without significant
orbital shrinkage.

\section*{Acknowledgments}

NI thanks NSERC Discovery and Canada Research Chairs Program. SJ thanks
the Chinese Academy of Sciences and National Science Foundation of
China, project numbers 11250110055 and 11350110324. 
PhP thanks the Umezawa Fund for hospitality.
This work was supported in part by the National Science Foundation under
Grant No. NSF PHY11-25915, Grant No. PHYS-1066293 and the hospitality of
the Aspen Center for Physics.

\bibliographystyle{mn2e} 
\bibliography{comenv}

\begin{thebibliography}{}

\bibitem[\protect\citeauthoryear{{De Marco}, {Passy}, {Moe}, {Herwig}, {Mac
  Low} \& {Paxton}}{{De Marco} et~al.}{2011}]{2011MNRAS.411.2277D}
{De Marco} O.,  {Passy} J.-C.,  {Moe} M.,  {Herwig} F.,  {Mac Low} M.-M.,
  {Paxton} B.,  2011, \mnras, 411, 2277

\bibitem[\protect\citeauthoryear{{Deloye} \& {Taam}}{{Deloye} \&
  {Taam}}{2010}]{DeloyeTaam10}
{Deloye} C.~J.,  {Taam} R.~E.,  2010, \apjl, 719, L28

\bibitem[\protect\citeauthoryear{{Fox} \& {Wood}}{{Fox} \&
  {Wood}}{1982}]{Fox+Wood1982}
{Fox} M.~W.,  {Wood} P.~R.,  1982, \apj, 259, 198

\bibitem[\protect\citeauthoryear{{Han}, {Podsiadlowski} \& {Eggleton}}{{Han}
  et~al.}{1994}]{Han+1994}
{Han} Z.,  {Podsiadlowski} P.,    {Eggleton} P.~P.,  1994, \mnras, 270, 121

\bibitem[\protect\citeauthoryear{{Han}, {Podsiadlowski} \& {Eggleton}}{{Han}
  et~al.}{1995}]{Han+1995}
{Han} Z.,  {Podsiadlowski} P.,    {Eggleton} P.~P.,  1995, \mnras, 272, 800

\bibitem[\protect\citeauthoryear{{Han}, {Podsiadlowski}, {Maxted}, {Marsh} \&
  {Ivanova}}{{Han} et~al.}{2002}]{Han+2002}
{Han} Z.,  {Podsiadlowski} P.,  {Maxted} P.~F.~L.,  {Marsh} T.~R.,    {Ivanova}
  N.,  2002, \mnras, 336, 449

\bibitem[\protect\citeauthoryear{{Ivanova}}{{Ivanova}}{2002}]{IvanovaThesis}
{Ivanova} N.,  2002, DPhil Thesis

\bibitem[\protect\citeauthoryear{{Ivanova}}{{Ivanova}}{2011}]{Ivanova11}
{Ivanova} N.,  2011, \apj, 730, 76

\bibitem[\protect\citeauthoryear{{Ivanova} \& {Chaichenets}}{{Ivanova} \&
  {Chaichenets}}{2011}]{Ivach11}
{Ivanova} N.,  {Chaichenets} S.,  2011, \apjl, 731, L36+

\bibitem[\protect\citeauthoryear{{Ivanova}, {Justham}, {Avendano Nandez} \&
  {Lombardi}}{{Ivanova} et~al.}{2013}]{Ivanova+2013Science}
{Ivanova} N.,  {Justham} S.,  {Avendano Nandez} J.~L.,    {Lombardi} J.~C.,
  2013, Science, 339, 433

\bibitem[\protect\citeauthoryear{{Ivanova}, {Justham}, {Chen}, {De Marco},
  {Fryer}, {Gaburov}, {Ge}, {Glebbeek}, {Han}, {Li}, {Lu}, {Marsh},
  {Podsiadlowski}, {Potter}, {Soker}, {Taam}, {Tauris}, {van den Heuvel} \&
  {Webbink}}{{Ivanova} et~al.}{2013}]{Ivanova+2013Review}
{Ivanova} N.,  {Justham} S.,  {Chen} X.,  {De Marco} O.,  {Fryer} C.~L.,
  {Gaburov} E.,  {Ge} H.,  {Glebbeek} E.,  {Han} Z.,  {Li} X.-D.,  {Lu} G.,
  {Marsh} T.,  {Podsiadlowski} P.,  {Potter} A.,  {Soker} N.,  {Taam} R.,
  {Tauris} T.~M.,  {van den Heuvel} E.~P.~J.,    {Webbink} R.~F.,  2013, \aapr,
  21, 59

\bibitem[\protect\citeauthoryear{{Ivanova} \& {Taam}}{{Ivanova} \&
  {Taam}}{2004}]{2004ApJ...601.1058I}
{Ivanova} N.,  {Taam} R.~E.,  2004, \apj, 601, 1058

\bibitem[\protect\citeauthoryear{{Kutter} \& {Sparks}}{{Kutter} \&
  {Sparks}}{1972}]{Kutter+Sparks1972}
{Kutter} G.~S.,  {Sparks} W.~M.,  1972, \apj, 175, 407

\bibitem[\protect\citeauthoryear{{Kutter} \& {Sparks}}{{Kutter} \&
  {Sparks}}{1974}]{Kutter+Sparks1974}
{Kutter} G.~S.,  {Sparks} W.~M.,  1974, \apj, 192, 447

\bibitem[\protect\citeauthoryear{{Ledoux}}{{Ledoux}}{1945}]{Ledoux1945}
{Ledoux} P.,  1945, \apj, 102, 143

\bibitem[\protect\citeauthoryear{{Lobel}}{{Lobel}}{2001}]{Lobel2001}
{Lobel} A.,  2001, \apj, 558, 780

\bibitem[\protect\citeauthoryear{{Meyer} \& {Meyer-Hofmeister}}{{Meyer} \&
  {Meyer-Hofmeister}}{1979}]{mmh79}
{Meyer} F.,  {Meyer-Hofmeister} E.,  1979, \aap, 78, 167

\bibitem[\protect\citeauthoryear{{Paczy{\'n}ski}}{{Paczy{\'n}ski}}{1969}]{1969AcA....19....1P}
{Paczy{\'n}ski} B.,  1969, \actaa, 19, 1

\bibitem[\protect\citeauthoryear{{Paczynski}}{{Paczynski}}{1976}]{Pa76}
{Paczynski} B.,  1976, in {P.~Eggleton, S.~Mitton, \& J.~Whelan} ed., Structure
  and Evolution of Close Binary Systems Vol.~73 of IAU Symposium, {Common
  Envelope Binaries}.
p.~75

\bibitem[\protect\citeauthoryear{{Paczy{\'n}ski} \&
  {Zi{\'o}{\l}kowski}}{{Paczy{\'n}ski} \& {Zi{\'o}{\l}kowski}}{1968}]{PZ1968}
{Paczy{\'n}ski} B.,  {Zi{\'o}{\l}kowski} J.,  1968, \actaa, 18, 255

\bibitem[\protect\citeauthoryear{{Passy}, {De Marco}, {Fryer}, {Herwig},
  {Diehl}, {Oishi}, {Mac Low}, {Bryan} \& {Rockefeller}}{{Passy}
  et~al.}{2012}]{Passy11}
{Passy} J.-C.,  {De Marco} O.,  {Fryer} C.~L.,  {Herwig} F.,  {Diehl} S.,
  {Oishi} J.~S.,  {Mac Low} M.-M.,  {Bryan} G.~L.,    {Rockefeller} G.,  2012,
  \apj, 744, 52

\bibitem[\protect\citeauthoryear{{Podsiadlowski}}{{Podsiadlowski}}{2001}]{podsi01}
{Podsiadlowski} P.,  2001, in {P.~Podsiadlowski, S.~Rappaport, A.~R.~King,
  F.~D'Antona, \& L.~Burderi } ed., Evolution of Binary and Multiple Star
  Systems Vol.~229 of Astronomical Society of the Pacific Conference Series,
  {Common-Envelope Evolution and Stellar Mergers}.
pp 239--+

\bibitem[\protect\citeauthoryear{{Quataert} \& {Shiode}}{{Quataert} \&
  {Shiode}}{2012}]{2012MNRAS.423L..92Q}
{Quataert} E.,  {Shiode} J.,  2012, \mnras, 423, L92

\bibitem[\protect\citeauthoryear{{Ricker} \& {Taam}}{{Ricker} \&
  {Taam}}{2008}]{Ric08}
{Ricker} P.~M.,  {Taam} R.~E.,  2008, \apjl, 672, L41

\bibitem[\protect\citeauthoryear{{Ricker} \& {Taam}}{{Ricker} \&
  {Taam}}{2012}]{Ric2012}
{Ricker} P.~M.,  {Taam} R.~E.,  2012, \apj, 746, 74

\bibitem[\protect\citeauthoryear{{Shiode} \& {Quataert}}{{Shiode} \&
  {Quataert}}{2014}]{2014ApJ...780...96S}
{Shiode} J.~H.,  {Quataert} E.,  2014, \apj, 780, 96

\bibitem[\protect\citeauthoryear{{Soker}}{{Soker}}{2008}]{2008ApJ...674L..49S}
{Soker} N.,  2008, \apjl, 674, L49

\bibitem[\protect\citeauthoryear{{Sparks} \& {Kutter}}{{Sparks} \&
  {Kutter}}{1972}]{Sparks+Kutter1972}
{Sparks} W.~M.,  {Kutter} G.~S.,  1972, \apj, 175, 707

\bibitem[\protect\citeauthoryear{{Stothers}}{{Stothers}}{1999}]{Stothers1999}
{Stothers} R.~B.,  1999, \mnras, 305, 365

\bibitem[\protect\citeauthoryear{{Tauris} \& {Dewi}}{{Tauris} \&
  {Dewi}}{2001}]{TaurisDewi01}
{Tauris} T.~M.,  {Dewi} J.~D.~M.,  2001, \aap, 369, 170

\bibitem[\protect\citeauthoryear{{Tuchman}, {Sack} \& {Barkat}}{{Tuchman}
  et~al.}{1978}]{Tuchman+1978}
{Tuchman} Y.,  {Sack} N.,    {Barkat} Z.,  1978, \apj, 219, 183

\bibitem[\protect\citeauthoryear{{Tuchman}, {Sack} \& {Barkat}}{{Tuchman}
  et~al.}{1979}]{Tuchman+1979}
{Tuchman} Y.,  {Sack} N.,    {Barkat} Z.,  1979, \apj, 234, 217

\bibitem[\protect\citeauthoryear{{Wagenhuber} \& {Weiss}}{{Wagenhuber} \&
  {Weiss}}{1994}]{W+W1994}
{Wagenhuber} J.,  {Weiss} A.,  1994, \aap, 290, 807

\bibitem[\protect\citeauthoryear{{Webbink}}{{Webbink}}{2008}]{Webbink08}
{Webbink} R.~F.,  2008, in {E.~F.~Milone, D.~A.~Leahy, \& D.~W.~Hobill} ed.,
  "Short-Period Binary Stars: Observations, Analyses, and Results" Vol.~352 of
  Astrophysics and Space Science Library, {Common Envelope Evolution Redux}.
pp 233--+

\bibitem[\protect\citeauthoryear{{Wood}}{{Wood}}{1974}]{1974ApJ...190..609W}
{Wood} P.~R.,  1974, \apj, 190, 609

\end{thebibliography}

\end{document}